\documentclass[amsmath,amssymb,aps,prl,reprint,showpacs,twocolumn,floats,superscriptaddress]{revtex4-2}

\usepackage{braket}
\usepackage[table,xcdraw]{xcolor}
\usepackage{graphicx}
\usepackage{amssymb}
\usepackage{amsmath}
\usepackage{hyperref}
\usepackage{footnote}
\usepackage{color}
\usepackage{multirow}
\usepackage{enumerate}
\usepackage{siunitx}
\usepackage{soul}
\usepackage{subfigure}
\pdfoutput=1 
\usepackage[utf8]{inputenc}
\makeatletter
\def\@fnsymbol#1{\ensuremath{\ifcase#1\or \dagger\or \ddagger\or  \mathsection\or
  \mathparagraph\or *\or \|\or **\or \dagger\dagger
   \or \ddagger\ddagger \else\@ctrerr\fi}}
\makeatother

\begin{document}
\title{Trade off-Free Entanglement Stabilization in a Superconducting Qutrit-Qubit System}
\author{T.~Brown}
\affiliation{Department of Physics and Applied Physics, University of Massachusetts, Lowell, MA 01854, USA}
\affiliation{Quantum Engineering and Computing, Raytheon BBN Technologies, Cambridge, MA 02138, USA}
\author{E.~Doucet}
\affiliation{Department of Physics and Applied Physics, University of Massachusetts, Lowell, MA 01854, USA}
\author{D.~Rist\`{e}}
\thanks{Present address: Keysight Technologies, Cambridge, MA 02139, USA}
\affiliation{Quantum Engineering and Computing, Raytheon BBN Technologies, Cambridge, MA 02138, USA}
\author{G.~Ribeill}
\affiliation{Quantum Engineering and Computing, Raytheon BBN Technologies, Cambridge, MA 02138, USA}
\author{K.~Cicak}
\affiliation{National Institute of Standards and Technology, 325 Broadway, Boulder, CO 80305, USA}
\author{J.~Aumentado}
\affiliation{National Institute of Standards and Technology, 325 Broadway, Boulder, CO 80305, USA}
\author{R.~Simmonds}
\affiliation{National Institute of Standards and Technology, 325 Broadway, Boulder, CO 80305, USA}
\author{L.~Govia}
\affiliation{Quantum Engineering and Computing, Raytheon BBN Technologies, Cambridge, MA 02138, USA}
\author{A.~Kamal}
\email{archana\_kamal@uml.edu}
\affiliation{Department of Physics and Applied Physics, University of Massachusetts, Lowell, MA 01854, USA}
\author{L.~Ranzani*}
\email{leonardo.ranzani@raytheon.com}
\affiliation{Quantum Engineering and Computing, Raytheon BBN Technologies, Cambridge, MA 02138, USA}
\begin{abstract}
Quantum reservoir engineering is a powerful framework for autonomous quantum state preparation and error correction. However, traditional approaches to reservoir engineering are hindered by unavoidable coherent leakage out of the target state, which imposes an inherent trade off between achievable steady-state state fidelity and stabilization rate. In this work we demonstrate a protocol that achieves trade off-free Bell state stabilization in a qutrit-qubit system realized on a circuit-QED platform. We accomplish this by creating a purely dissipative channel for population transfer into the target state, mediated by strong parametric interactions coupling the second-excited state of a superconducting transmon and the engineered bath resonator. Our scheme achieves a state preparation fidelity of $84\%$ with a stabilization time constant of $339\,$ns, leading to the lowest error-time product reported in solid-state quantum information platforms to date. 
\end{abstract}
\maketitle
%
%
\section{Introduction}
%
%
Entanglement is a fundamental property of quantum systems and is essential to achieve quantum advantage in almost any application of quantum information processing, such as sensing~\citep{degen2017quantum}, communication~\citep{gisin2007quantum} and computing~\citep{arute2019quantum}. Typically, entanglement is created by applying a sequence of single and two-qubit unitaries; however the resulting states are subject to decoherence caused by coupling to the surrounding environment~\citep{blais2007quantum}. In the absence of active error correction~\citep{fowler2012surface,riste2013deterministic,sayrin2011real,riste2012feedback,andersen2019entanglement}, decoherence limits the circuit depth and, consequently, the size of the entangled state that can be produced. Moreover, this approach is sensitive to state preparation and measurement (SPAM) errors which accumulate as the complexity and size of the quantum system increases. An attractive alternative for quantum state preparation is quantum reservoir engineering, where a quantum system is steered to a desired entangled state by coupling it to an auxiliary system (`engineered reservoir') that induces strong, non-local dissipation on the target system~\cite{Doucet2020,gertler2021protecting,kimchi2016stabilizing,lu2017universal,shankar2013autonomously,murch2012cavity,lin2013dissipative,poyatos1996quantum,pichler2015quantum,sarlette2011stabilization,didier2018remote}. In addition to being immune to initialization errors, the target state remains stabilized for times much longer than the coherence time of the individual qubits, ensuring the entangled state is always available on demand. 
\par
Though there have been several demonstrations of dissipative stabilization in diverse quantum information platforms, such as superconducting qubits~\cite{lu2017universal,shankar2013autonomously,leghtas2015confining,grimm2020stabilization,murch2012cavity}, trapped ions~\cite{lin2013dissipative,cole2021dissipative,horn2018quantum}, atomic systems~\cite{krauter2011entanglement}, and NV centers~\cite{li2012dissipative}, almost all reported schemes have been hindered by unavoidable coherent leakage out of the target state that cannot be suppressed without also reducing the repumping rate into the desired state. This issue, in fact, leads to a trade off in reservoir engineering: the product of minimum steady-state error ($\varepsilon_{\infty}$) and stabilization time ($\tau$) is a constant that is independent of the engineered dissipation rate, implying that \emph{perfect} entanglement stabilization cannot be achieved at rates faster than the uncontrolled dissipation rates. This severely limits the prospects of reservoir engineering both in terms of (i) usability with regard to implementation in systems with strong local (uncontrolled) decoherence, which ironically stand to gain most from such stabilization techniques, and (ii) scalability with regard to state preparation in large quantum networks, where it becomes increasingly harder for the stabilization rate to beat the cumulative local decoherence, which scales (at least) linearly with system size \cite{SchlosshauerRMP2005,Xu2019, Berke2020}. 
%
%
%
\par
Nonetheless, as shown by our recent work~\cite{Doucet2020} such a trade off is not a fundamental limitation of autonomous state stabilization and, in fact, it is a consequence of driven-dissipative schemes that transfer population into a target state at a rate limited by a drive strength which needs to remains weak (or `perturbative') as compared to the dressed linewidth to maintain resonant pumping. The concurrent scaling of terminal fidelity and stabilization rate can be achieved instead by engineering stabilization protocols that realize desired entangled states as an eigenstate of the drive Hamiltonian that is \emph{simultaneously} also protected from dissipation (\textit{i.e.} it is a dark state of the dynamics). Henceforth, we term such stabilization protocols `exact', as they stabilize the target state with unit fidelity in the absence of local decoherence.
\begin{figure*}[t!]
    \centering
    \includegraphics[width=\textwidth]{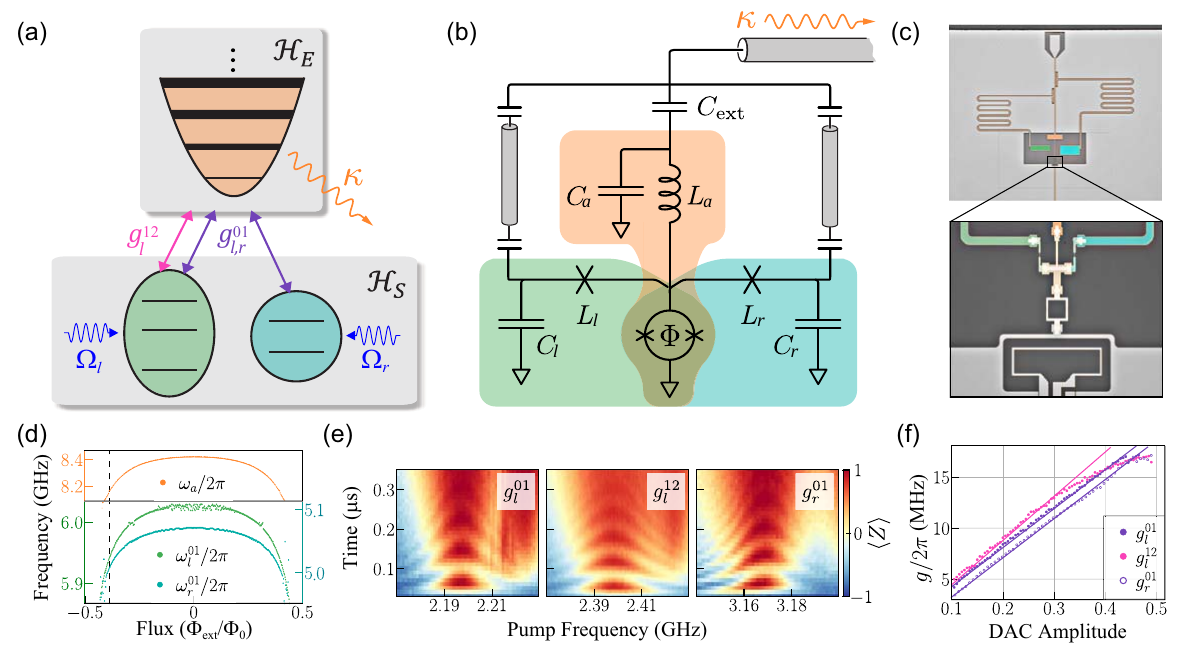}
    \caption{(a) Schematic diagram showing a lossy resonator with linewidth $\kappa$ coupled to a qutrit-qubit system via three parametric drives. Rabi drives resonant with the 0\,-1 transition are also applied to both the qutrit and the qubit. (b) Circuit realization of the scheme in (a) showing two superconducting transmons (green and turquoise) and central resonator (orange), as well as the dedicated readout resonators (grey). The transmons and central resonator share a SQUID that implements parametric couplings ($g_{k}^{n,n+1}$). (c) Optical micrographs of the experimental device layout (left) and a magnified view of the junctions and the SQUID coupler (right). The external bias line used to pump the SQUID can be seen at the bottom of the device. The resonator consists of an array of 10 Josephson junctions, each having a critical current $I_{ac} \approx 540\,$nA, in series with a fixed capacitor. (d) Measured 0\,-1 transition frequency for each transmon and resonator center frequency as a function of flux through the SQUID loop. The operating flux bias is indicated with a dashed-black line. (e) Time-domain parametric swaps measured as a function of pump frequency for each of the three parametric drives in Eq.~(\ref{eqn2}). From left to right, the transmons are initialized in $|10\rangle$, $|20\rangle$, and $|01\rangle$ respectively. (f) Parametric coupling rates measured as a function of pump amplitude. Solid lines are linear fits to the data; the nonlinear response of $g_l^{12}$ at higher drive amplitude is due to enhanced mixer saturation at the corresponding IF frequency of \SI{150}{\MHz}, as compared to the other drives, which use \SI{50}{\MHz}. }
    \label{fig:fig1}
\end{figure*}
\par
In this work, we present the first implementation of an exact Bell-state stabilization protocol in a superconducting circuit-QED system comprising two transmons parametrically coupled to a common lossy resonator that acts as an engineered reservoir. We engineer a purely dissipative channel for population transfer into the target Bell state via parametric coupling to the third level of the transmon, and without any direct coherent coupling into or out of it, making it an eigenstate of the drive Hamiltonian and also a dark state of the engineered dissipation. Our scheme attains a steady-state fidelity of $84\%$ with a time-constant of \SI{339}{\ns} achieving the smallest error-time product $\varepsilon_{\infty}\tau \simeq \SI{54}{\ns}$ reported in quantum information platforms to date. Furthermore we verify that the steady state error and preparation time are linearly correlated, confirming the trade off-free behaviour. Notably, the reported protocol is the minimal instance of exact stabilization physics that employs only unconditional driving and linear (engineered) dissipation. 

\subsection{System and Effective Hamiltonian}
%
Fig.~\ref{fig:fig1}(a) depicts the general scheme, in which a qutrit-qubit system is coupled to a lossy resonator using bilinear parametric interactions. Our circuit-QED implementation in Fig.~\ref{fig:fig1}(b) consists of two transmon qubits coupled to a superconducting resonator. The parametric interactions are realized by grounding the resonator and the transmon junctions through a shared Superconducting Quantum Interference Device (SQUID) loop, which acts as a flux-tunable inductor. Through sinusoidal modulation of the flux through the SQUID loop, ${\Phi(t) = \Phi_{\rm ext}+ \sum_{j}\Phi_{j}\cos(\omega_{j}t+\phi_{j})}$, pairwise couplings can be activated between any pair of elements via the choice of the pump frequency $\omega_{j}$. We show the layout of the experimental device in Fig.~\ref{fig:fig1}(c), with detailed parameters listed in~\citep{Supplement}. 
%
%
%
\par
We simultaneously activate the parametric couplings depicted in Fig.~\ref{fig:fig1}(a) by flux-pumping the SQUID at the sideband frequencies $\omega_a \pm \omega_k^{n,n+1}$, corresponding to the desired transition frequencies of the transmon $k=l,r$. Specifically, we pump the two red sideband frequencies $\omega_{a} - \omega_{l}^{01}$ and $\omega_{a} - \omega_{l}^{12}$ corresponding to the 0\,-1 and 1-2 transitions for transmon $l$, and the red sideband at $\omega_{a} - \omega_{r}^{01}$ corresponding to the 0\,-1 transition of transmon $r$. In conjunction with Rabi drives on the 0\,-1 transitions of each qubit, this leads to an effective interaction Hamiltonian of the form~\cite{Supplement}:
\begin{eqnarray}
    H_{\textrm{I}} &=& a^\dagger \left(\frac{g_{l}^{12}}{2} e^{i \phi_{l}^{12}} |1 \rangle_l \langle 2|+\sum_{k=l,r} \frac{g_{k}^{01}}{2} e^{i \phi_{k}^{01}} |0 \rangle_k \langle 1| \right) \nonumber\\
    & & \quad + \sum_{k \in \{l,r\}}\frac{\Omega_k^{01}}{2} e^{i \theta_{k}} |0 \rangle_k \langle 1| + h.c.,
    \label{eqn2}
\end{eqnarray}
where we have moved to a frame defined w.r.t. the free Hamiltonian and discarded off-resonant counter-rotating terms~\cite{Supplement}. The parametric couplings shuttle excitations between the transmon levels and the lossy resonator~\citep{zakka2011quantum,sirois2015coherent,lu2017universal}, leading to an engineered quasi-local dissipator $\mathcal{D}\left[L_{\rm eff}\right]$ acting on the transmons with 
\begin{eqnarray}
    L_{\textrm{eff}} = c_{l}^{12} |1 \rangle_l \langle 2|+\sum_{k\in \{l,r\}} c_{k}^{01} |0 \rangle_k \langle 1|.
    \label{eqn3}
\end{eqnarray}
Here the coefficients $c_{k}^{n,n+1}$ are functions of the resonator decay rate $\kappa$ and the corresponding parametric pump amplitudes ($g_{k}^{n,n+1}$) and phases ($\phi_{k}^{n,n+1}$). In particular, the phases $\phi_{k}^{n,n+1}$ are needed to perform coherent control of the stabilized state, as explained later. We choose our coupling rates so that the target Bell state $|\psi\rangle$ is an eigenstate of the Hamiltonian in Eq.~(\ref{eqn2}) and satisfies $L_{\rm eff}|\psi\rangle=0$~\citep{Doucet2020,zoller2008preparation}. For example, in order to prepare the state $|\psi\rangle=(1/\sqrt{2})(|01\rangle+e^{i\phi}|10\rangle)$ we set $g_{l}^{01}=-e^{i\phi}g_{r}^{01}$ and $\Omega_l^{01}=-e^{i\phi}\Omega_r^{01}$.  
\par
In our experiment, we bias the SQUID coupler at $\Phi_{\rm ext}=-0.39\Phi_0$ corresponding to transmon 0\,-1 transition frequencies of $\omega^{01}_l=2\pi\times \SI{5.928}{\GHz}$ and $\omega^{01}_r=2\pi\times \SI{4.993}{\GHz}$ 
and center frequency of the resonator $\omega_a=2\pi\times \SI{8.124}{\GHz}$ [Fig.~\ref{fig:fig1}(d)]. 
We characterize the parametric drive amplitudes $g_{l,r}^{01}$ by initializing each transmon in state $|1\rangle_{l,r}$ and subsequently turning on the respective red-sideband $\omega_{a} - \omega_{l,r}^{01}$ to measure the coherent swaps with the resonator, see Fig.~\ref{fig:fig1}(e,f). Similarly, we measure $g_{l}^{12}$ by initializing the left transmon in the state $|2\rangle_{l}$ using a sequence of $\pi^{01}, \pi^{12}$ pulses, the first resonant with the $\omega_l^{01}$ and the second resonant with the $\omega_l^{12} = \omega_l^{01}+\alpha_l$, where the anharmonicity $\alpha_l=-2\pi\times\SI{198}{\MHz}$. We then measure the coherent swap between the states $|2\rangle_l|0\rangle_{a}$ and $|1\rangle_l|1\rangle_{a}$ under the action of red-sideband at $\omega_{a} - \omega_{l}^{12}$. In Fig.~\ref{fig:fig1}(e) we show an example of time-domain oscillations, indicating coherent swap with decay time approximately equal to $2/\kappa$, which is consistent with the hybridization between the transmon transitions and the lossy resonator. We fit the oscillations on resonance corresponding to each transition to a decaying sinusoid and extract the parametric coupling rate from the swap period $1/g$; we do this for different drive amplitudes and measure coupling rates up to $g=2\pi\times\SI{17.5}{\MHz}$ [Fig.~\ref{fig:fig1}(f)].   
\begin{figure}[t!]
    \centering
  \includegraphics[width=\columnwidth]{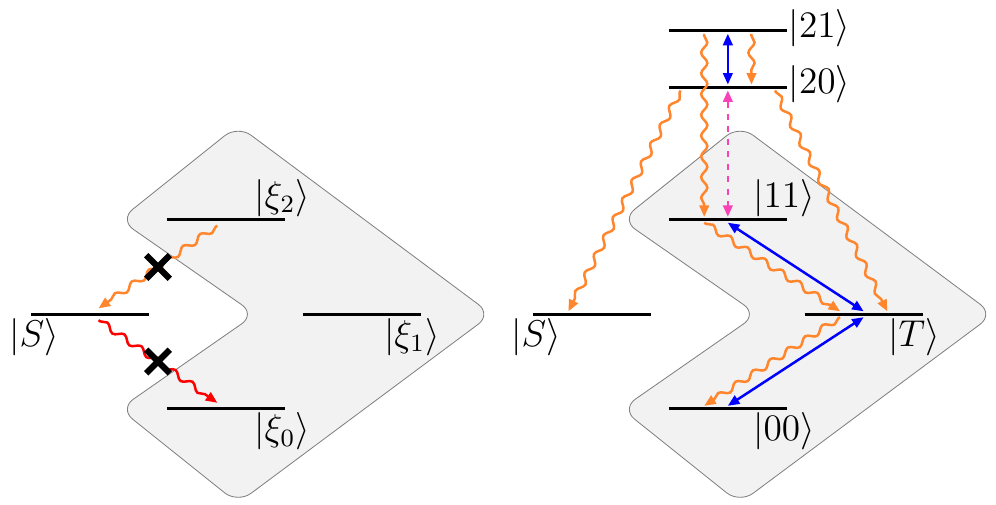}
    \caption{(a) Quasilocal dissipation engineered with red and blue sideband qubit-oscillator interactions: it is not possible to protect the target Bell state $|S\rangle$ from decay (red arrow) without also suppressing any repumping channel from the orthogonal subspace into the target state (orange arrow). (b) The effective action of our protocol on the reduced qutrit-qubit subspace. Note how the qutrit levels provide the triplet subspace an indirect path to decay into the singlet subspace.}
    \label{fig:fig2}
\end{figure}
\par
We emphasize that this qutrit-qubit scheme employs the \emph{minimal} set of interactions necessary for exact stabilization of a two-qubit maximally entangled state. Specifically, including one extra level --- here the second-excited state of the left transmon, enables achieving exact state stabilization using only quasi-local dissipation. We can see why by the following general argument. Let us assume without loss of generality that we want to stabilize the singlet state $|S\rangle=(1/\sqrt{2})(|01\rangle-|10\rangle)$. The most general jump operator restricted to a two-qubit (four-level) space that satisfies $L|S\rangle = 0$ is $L=c_{-}J^{-}+c_{+}J^{+}$ where $J^{\pm}$ are the total spin raising and lowering operators~\cite{Supplement}. Similarly, if $\left|S\right\rangle$ is an eigenstate of the system Hamiltonian, the latter commutes with the total spin $\left[H, J^2\right] = 0$. We conclude that since every generator commutes with $J^2$, the total spin is a conserved quantity and hence the qubits cannot be stabilized into $\left|S\right\rangle$ (corresponding to $J=0$) when initialized in a state corresponding to different total spin ($J=1$). Figure~\ref{fig:fig2}(a) illustrates the problem -- that in the process of engineering $|S\rangle$ as the steady state of dissipation, we completely decouple it from the rest of the Hilbert space. This ``no-go" argument can be extended to the stabilization of any Bell state, by redefining the spin operators up to a local unitary on either qubit.
\begin{figure*}[t!]
    \centering
    \includegraphics[width=\textwidth]{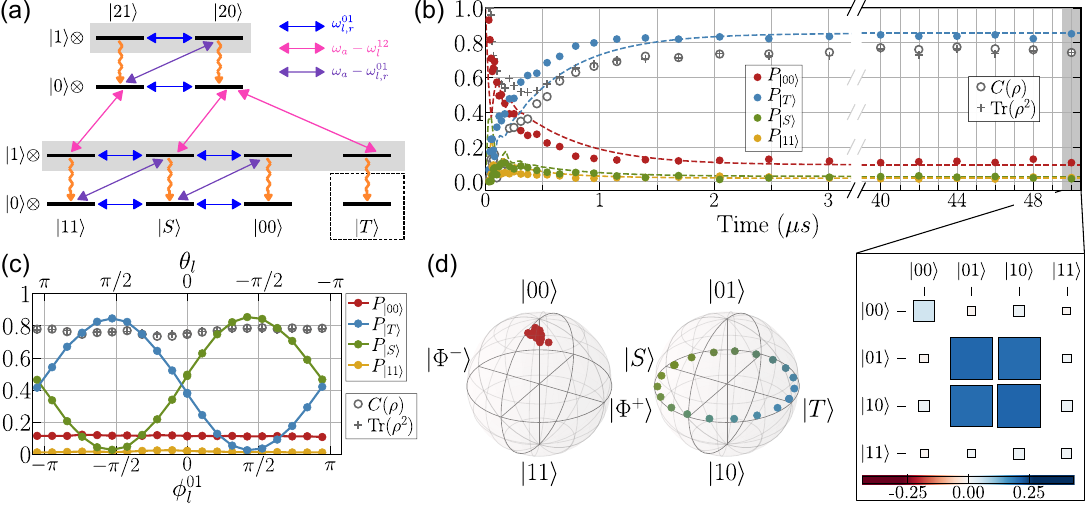}
    \caption{(a) Energy level diagram showing the action of $\Omega_{l,r}$ (blue), $g_{l,r}^{01}$ (purple), $g_{l,r}^{12}$  (magenta), and $\kappa$ (orange). Only the first two energy levels of the resonators are shown for clarity. (b) Stabilization trajectory for transmons initialized in $|00\rangle$  obtained for $\Omega_{l,r}=2\pi\times \SI{7.2}{\MHz}$, $g^{01}_{l,r}=2\pi\times \SI{7.5}{\MHz}$, $g^{12}_l=2\pi\times\SI{13.1}{\MHz}$, $\kappa=2\pi\times\SI{4.73}{\MHz}$, showing target state (here $|T\rangle$) fidelity of 84\% fidelity stabilized for \SI{50}{\micro\second}. The dashed line corresponds to trajectories obtained using master equation simulations. Density matrix at $t=\SI{50}{\micro\second}$, reconstructed using quantum state tomography, shows that most of steady-state error is accounted for by decay into $|00\rangle$. (c) Coherent control of the stabilized state, with simultaneous tuning of the phases of Rabi ($\Omega_l$) and parametric drives ($g_l^{01}$), maintaining $\phi_{l}^{01} + \theta_{l} = 0$. The populations are shown at a fixed time $t = \SI{2}{\micro\second}$ as a function of drive phase and remaining drive phases fixed. The average purity and concurrence over the $2\pi$ rotation are 77\%. (d) Two-qubit Bloch sphere representations for (normalized) projections of steady state in odd- and even-parity manifolds.}
    \label{fig:fig3}
\end{figure*}
\par
In the past~\cite{zoller2008preparation}, this issue has been bypassed by stabilizing an `approximate' Bell state, $|\psi\rangle = \mathcal{N}_{\delta}( |S\rangle + \delta |\xi\rangle)$ with $ L^{\dagger}|\xi\rangle \neq 0$, thus allowing the use of interactions which do not conserve the total spin. This approach, however, only enables a perturbative stabilization of Bell states, which results in a trade off between fidelity and time of the protocol~\cite{Supplement}. In contrast, the protocol presented here circumvents this problem by expanding the system Hilbert space to a \emph{qutrit}-qubit system. The sideband coupling to the second excited state of the qutrit ($g_l^{12}$) leads to a jump operator as indicated in Eq.~(\ref{eqn3}) that does not preserve the ``total spin'' on the qubit-qubit subspace but still supports $\left|S\right\rangle$ as an exact steady state. Figure~\ref{fig:fig2}(b) illustrates the mechanism of our protocol after adiabatic elimination of the oscillator \cite{CarmichaelVol2}. There is no direct coupling between the singlet and triplet subspaces, either dissipative or coherent. Instead, the additional levels included in the Hilbert space due to the second excited state of the qutrit mediate a pathway for population to decay from the triplet to the singlet subspace in a multi-step process.
\subsection{Stabilization Mechanism and Performance}
%
%
%
The stabilization procedure implemented by the Hamiltonian in Eq.~(\ref{eqn2}), in conjunction with the resonator decay ($\kappa$), is able to stabilize any odd-parity Bell state through appropriate choice of phases $\phi_{k}, \theta_{k}$. Fig.~\ref{fig:fig3}(a) shows the full energy level diagram, including the resonator levels, depicting the mechanism for stabilization of triplet state $|T\rangle=(1/\sqrt{2})(|01\rangle+|10\rangle)$ as an example. Here, the two 0\,-1 sideband drives are set out-of-phase (purple arrows) to selectively couple only the orthogonal state $|S\rangle$ to the even-parity states in the one-excitation manifold of the resonator. The out-of-phase Rabi drives (blue arrows) couple $|00\rangle$ (and $|11\rangle$) to $|S\rangle$ and prevent the system from being trapped in $|00\rangle|0\rangle_a$. Finally, the 1-2 sideband drive (magenta arrows) couples the second-excited state of the left transmon to the states in the one- and two-excitation manifolds of the resonator. The combined action of the three parametric and two direct drives pumps the population into $|T\rangle|1\rangle_{a}$, which then quickly decays to the target state $|T\rangle|0\rangle_{a}$. Crucially, no drive acts on $|T\rangle|0\rangle_{a}$ directly, making it an exact dark state of the driven-dissipative evolution as explained before.  
\par
For demonstrating stabilization, we initialize both transmons in their ground state since it requires no prior active preparation, simultaneously turn on all three parametric drives and two Rabi drives for a fixed time $t$, and finally perform two-qubit quantum state tomography to reconstruct the evolution of the two-qubit state as a function of $t$. We note here that stabilized state is independent of the initial state of the qubits; data for different initial states are discussed in~\cite{Supplement}. 
\begin{figure*}[t!]
\begin{minipage}[r]{0.52\textwidth}
\includegraphics[width=\textwidth]{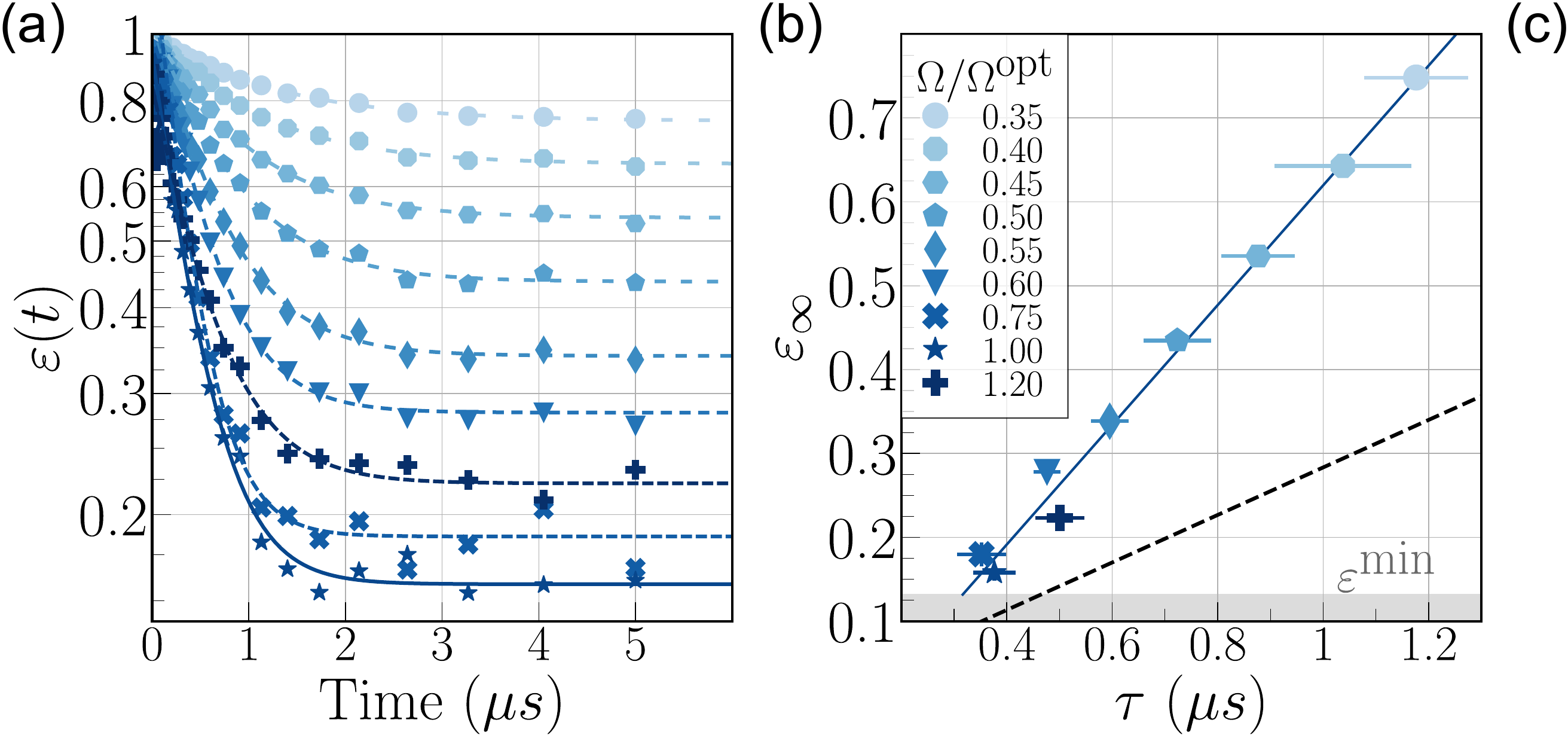}%
\end{minipage}%
\begin{minipage}[l]{0.49\textwidth}
\bgroup
\def\arraystretch{1.22}
\begin{tabular}[t]{c|cccc}
\noalign{\hrule height 1pt}
& Lin \cite{lin2013dissipative} & Shankar \cite{shankar2013autonomously} & K.S. \cite{kimchi2016stabilizing} & \textbf{This work}\\
& Ions & cQED & cQED & cQED 
\\
\noalign{\hrule}
\multirow{2}*{$T_2^*$} & $100 \si{\micro\second}$ \cite{T2ion} &  $8 \si{\micro\second}$ & $2.6 \si{\micro\second}$ & $5.6 \si{\micro\second}$ \\
& & $12 \si{\micro\second}$ & $3.0 \si{\micro\second}$ & $4.5 \si{\micro\second}$ 
\\
\noalign{\hrule}
$F_{\infty}^{\rm max}$ & 77\% & 67\%  & 71\% & 84\% \\
$\tau$ & $>1 \si{\milli\second}$ & $960 \si{\nano\second}$ & $760 \si{\nano\second}$ & $339 \si{\nano\second}$ 
\\
\noalign{\hrule}
$~\varepsilon_{\infty}^{\rm min} \tau~$ & $>200 \si{\micro\second}$ & $317 \si{\nano\second}$ & $220 \si{\nano\second}$ & $54\si{\nano\second}$ \\
$I_{e}$ ($T_{2}^{*}$) & 0.01 & 4.47 & 3.21 & 6.23 \\
\noalign{\hrule height 1pt}
\end{tabular}%
\egroup
\end{minipage}%
\caption{(a) Measurement of preparation error as a function of time, $\varepsilon(t)$, for different Rabi drive amplitudes of $\Omega_{l,r}^{01}$. (b) Parametric plot of the steady-state error $\varepsilon_{\infty}$ and the stabilization time constant $\tau$, obtained from fitting the data in (a) showing the expected linear scaling. The minimum $\varepsilon_{\infty}$ and $\tau$ are measured at $\Omega_{l,r}^{\rm opt}\approx 1.5\kappa$. The minimum error $\varepsilon_{\infty}^{\rm min}$ is close to theoretically achievable value for experimental $\kappa$ (shown as the grey floor) \cite{Supplement}. The dotted line is the error-time scaling obtained from a simulation of Eq.~(\ref{eqn2}) and has slope equal to $1/T_{B}$ with $T_{B} = \SI{3.4\,}{\micro\second}$, independent of drive amplitudes, in accordance with the semi-classical estimate discussed in the Methods section. The deviation of experimentally observed slope from $1/T_{B}$ can be quantitatively explained by including a residual drive detuning and imbalance in the sideband amplitude~\cite{Supplement}.
(c) Table comparing the dominant decoherence rates and performance metrics achieved in this work against previous implementations of dissipative Bell state stabilization. The quoted performance metrics are for continuous-wave (CW) driven and autonomous protocols, without any postselection, similar to the one in the present work \cite{LinNote}. The composite metrics $\varepsilon_\infty \tau$ and $I_e (x)$ provide a platform-agnostic means for comparing different stabilization protocols. \newline
}
\label{fig:scaling}
\end{figure*}
\par
We examine the stabilization time constant, $\tau$, and steady-state error, $\varepsilon_{\infty}$, from fitting the dynamical error for a given set of drive parameters as ${\varepsilon(t)=\varepsilon_{\infty} + \tilde{\varepsilon}\exp(-t/\tau)}$, where ${\varepsilon(t) = 1 - {\rm Tr}\{\rho(t)\mathbb{I}_{\rm res}\otimes |T\rangle\langle T|\}}$~\citep{Doucet2020}. With the Rabi drives tuned to their optimal coupling strength, the average stabilization trajectory displays an exponential behavior with a characteristic $1/e$ time of \SI{339}{\ns} as it approaches its steady state fidelity of 84(1)\%, found from the average and standard deviation of the data points between \SI{10}{\micro\second} and \SI{50}{\micro\second} [Fig.~\ref{fig:fig3}(b)]. The remaining population resides primarily in the ground state $|00\rangle$ (11\%), with small residuals in $|11\rangle$ (2\%) and the orthogonal Bell state $|S\rangle$ (2\%). We verified that the qubits remain in the steady state for as long as the pumps are on up to \SI{50}{\micro\second}, which is about 10$\times$ longer than the timescale set by the decoherence time of each qubit $T_2^{*}$~\cite{Supplement}. In Fig.~\ref{fig:fig3}(b) we also show a full master equation simulation of the system and a tomogram of the reconstructed two-qubit density matrix at $t = \SI{50}{\micro\second}$. The simulations, which are entirely based on independently measured device parameters, predict the correct target state fidelity within our measurement uncertainty. The measured convergence time is about 15\% faster than predicted by the theory for the measured drive amplitudes, which we attribute to residual drive detuning and imbalance, see also discussion on Fig.~\ref{fig:scaling} and~\cite{Supplement}. 
%
%
%
We separately characterize leakage out of the qubit manifold by measuring the population of the $|2\rangle_l$ state, which peaks at $t = \SI{150}{\nano\second}$ and then drops to less than 2.5\% at $t > \SI{1}{\micro\second}$. In the Supplement~\cite{Supplement} we provide details of obtaining a numerical bound on the error caused by this high-dimensional leakage ($< 1$\%) in the steady state reconstructed from two-qubit state tomography.
\par
By exploiting the tunable nature of parametric interactions, we realize \emph{in-situ} coherent control within a fixed parity manifold. For example, the stabilized Bell state can be rotated by tuning the phases of the 0\,-1 sidebands and Rabi drives while ensuring $\phi_{l}^{01}+\theta_{l}=\phi_{r}^{01}+\theta_{r}$. Such phase tuning allows for selection of any maximally entangled state while maintaining the purity, $P(\rho)$, and concurrence, $C(\rho)$, of the two-qubit state, as demonstrated in Fig.~\ref{fig:fig3}(c). Additionally, it is possible to continuously move along the longitude from $|01\rangle$ to $|10\rangle$ by changing the ratio between $g^{01}_{l}/g^{01}_{r}$ and $\Omega_{l}/\Omega_{r}$. The post-selected two-qubit state [Fig.~\ref{fig:fig3}(d)] shows an average purity of 95\% and 80\% in odd- and even-parity manifolds respectively. Using these numbers to model the full two-qubit state as a probabilistic mixture of even and odd parity subspaces, $\rho_{4\times 4} = x \rho_{\rm even} \oplus (1-x) \rho_{\rm odd}$, allows us to extract an improved post-selected fidelity of around 90\% for the odd-parity Bell state. 
\par
The main distinctive feature of our stabilization protocol is the concurrent scaling of preparation time and steady-state error. This is confirmed by the data presented in Fig.~\ref{fig:scaling} which shows a linear relationship between $\tau$ and $\varepsilon_{\infty}$; both decrease as the Rabi drive strength is increased, reaching a minimum near the optimal ${\Omega_{l,r}^{\rm opt} \approx 1.5\kappa}$. Simulations of both the full and reduced system (the latter obtained via adiabatic elimination of the resonator~\cite{Supplement}) confirm the linear error-time scaling, with the slope determined almost entirely by the total decoherence rate of the stabilized Bell state $1/T_{B} = \sum_{k\in l,r} (\gamma_{k}^{01}+\gamma_{k}^{11})/2$, i.e. $\varepsilon_{\infty}/\tau \approx 1/T_{B}$. Here $\gamma_{k}^{01}$ is the 0-1 relaxation rate and $\gamma_{k}^{11}$ is the pure dephasing rate. The linear relationship between $\varepsilon_{\infty}$ and $\tau$ indicates that the steady-state error is due to competition between engineered dissipation and local decoherence rates rather than any coherent error process. 
\par
The line corresponding to the experimentally measured stabilization error and time has a steeper slope than $1/T_{B}$ [c.f.~Fig.~\ref{fig:scaling}(b)]. Our simulations agree with the measured data when we include the effect of parametric crosstalk-induced drive detuning and amplitude imbalance, both of which lead to coherent leakage out of the target state~\citep{Doucet2020}. The fitted detunings (about 400 kHz on average) are a small fraction ($\leq 3\%$) of the measured power-dependent frequency shift of the right transmon (18 MHz), (see~\cite{Supplement} for details). We stress here that this coherent leakage remains weak compared to the engineered dissipation strength and, therefore, does not alter the expected error-time linear scaling.
\par
To assess the performance reported here against previous experiments we compare the product of $\varepsilon_\infty^{\rm min}$ and $\tau$, which can be thought of as an ``inverse gain-bandwidth product'' for state preparation. Our experiment yields $\varepsilon_\infty^{\rm min} \tau = \SI{54}{\nano\second}$, which is 5$\times$-6$\times$ lower than previous implementations of continuous-wave state stabilization in superconducting circuits [Table~\ref{fig:scaling}(c)]. We also propose an `information-theoretic' metric that allows to compute the upper bound on error-free output information generated by a multiple repetitions of a stabilization protocol. To this end, we model a stabilization cycle as a noisy binomial channel with the maximum success probability set by $p = 1- \varepsilon_{\infty}^{\rm min}$ and the number of uses set by the ratio $n_{c}=T_{c}/\tau$, where $T_{c}^{-1}$ is the repetition rate of the experiment. Assuming the output to be a continuous normal distribution, the expressions for maximum output entropy $H_{\rm max} = {\rm log}_{2} (2\pi e (S_{e}+N_{e}))$ and conditional entropy $H_{\rm error} = {\rm log}_{2} (2\pi e N_{e})$, with $S_{e}$ and $N_{e}$ being the root-mean-square signal and noise of the distribution, lead to the following expression for maximum value of entanglement efficiency $\mathcal{E}_e$ for a given scheme, 
\begin{eqnarray}
\mathcal{E}_e^{\rm max} (T_{c})
&=& \frac{\tau}{T_{c}} {\rm log}_{2}
\left(1+\frac{(1-\varepsilon_{\infty}^{\rm min})}{\varepsilon_{\infty}^{\rm min}} \frac{T_{c}}{\tau}\right).
\label{eqn:ebit}
\end{eqnarray}%
This can be understood as the maximum ``rate" at which the protocol can encode a continuous stream of bit-pairs into e-bits~\citep{eq4note}. For short enough repetition time, instead of using the minimum error, time dependent error can be used to calculate the relevant efficiency. Last row of Table~\ref{fig:scaling}(c) quotes the upper bound on the average information (here number of e-bits) generation capacity, $I_e (x) = (x/\tau) \mathcal{E}_{e}(x)$, at a fixed time set by the decoherence rates of different platforms.
%
\subsection{Conclusion}
%
In this work we have demonstrated an autonomous scheme which implements fast \emph{and} high-fidelity Bell state stabilization in a qutrit-qubit system. Use of parametric system-bath interactions allows operating the protocol with strong drive strengths (`engineered' dissipation) -- a regime which has hitherto remained inaccessible to reservoir engineering protocols based on resonant interactions. We verify that the preparation error scales linearly with the stabilization time constant, achieving a minimum error-time product of $\varepsilon_{\infty}^{\rm min} \tau = 54$~ns for optimal drive strengths. The non-monotonic variation of error-time product with drive amplitude results from a simultaneous minimization of drive-dominated and dissipation-dominated errors~\citep{Doucet2020}, highlighting a crucial principle for design of reservoir engineering schemes. Further, we implemented continuous-wave coherent control and \emph{in-situ} target state selection leveraging the phase tunability of parametric system-bath interactions.
\par
Further improvements of the proposed scheme using simple design variations, such as using parametric qubit-qubit drives instead of direct drives \cite{Supplement} and a moderate increase in resonator linewidth possibly coupled with the addition of a Purcell filter~\cite{reed2010fast}, can lead to 8-10\% higher fidelity with current hardware.
Since most of the state preparation error is due to the residual ground state population, a straightforward improvement in fidelity is achievable by using the center resonator to herald based on the state parity~\cite{riste2013deterministic,shankar2013autonomously}. The design principles underlying this work provide a novel addition to the parametric toolbox for quantum control in systems with strong light-matter interactions and can be readily extended for stabilization of multi-partite entangled states in large quantum networks.
%
%
%
%
\section{Methods}
%

\subsection{Simulation Approach}
We simulated our scheme using the following Lindblad master equation:

\begin{eqnarray*}
\dot{\rho} &=& -i[H_{I},\rho] + \kappa\mathcal{D}[a]\rho \\
& & + \sum_{k\in \{l,r\}}\left( \gamma_{k}^{01} \mathcal{D}[|0\rangle_{k}\langle 1|] + \gamma_{k}^{12} \mathcal{D}[|1\rangle_{k}\langle 2|] \right.\\
 & & \left. \hspace{1.2cm}+ 2\gamma_{k}^{11}\mathcal{D}[|1\rangle_{k}\langle1|]+ 2\gamma_{k}^{22}\mathcal{D}[|2\rangle_{k}\langle2|] \right)\rho, 
\end{eqnarray*}
where $\mathcal{D}[o]\rho = o\rho o^{\dagger} - \frac{1}{2}\{o^{\dagger}o,\rho\}$ and $H_I$ denotes the interaction Hamiltonian in Eq.~(\ref{eqn2}). In order to simulate pump amplitude-dependent shifts, we also include Hamiltonian terms of the form $\sum_{k=l,r}\delta_{k}^{01}|1\rangle_{k}\langle1|$ and $\delta_{a}a^\dag a$ that describe qubit and resonator detunings respectively. The measured relaxation rates are $\gamma_{l,r}^{01}$ and $\gamma_{l,r}^{12}$. The pure dephasing rates for $|0\rangle+|1\rangle$ and $|0\rangle+|2\rangle$ are $\gamma_{l,r}^{11}$ and $\gamma_{l,r}^{22}$ respectively. We have assumed that relaxation is a sequential process $2 \rightarrow 1 \rightarrow 0$ and cross-dephasing terms can be ignored ~\cite{peterer2015decay}. The latter approximation is justified since dephasing in our device is primarily due to thermal photons in the resonator. Moreover, we neglect the 0-2 decay process as it is a forbidden transition as per selection rules of the transmon. Detailed list of experimental parameters used for performing master equation simulations is included in~\cite{Supplement}.
For performing simulations, the absolute and relative tolerances of the 12th-order Adams-Moulton solver in QuTiP \cite{qutip} are each set to $10^{-12}$. Further, we truncate the Hilbert space corresponding to a maximum photon number $n=6$ in the resonator, beyond which we do not observe any appreciable change in the simulated Liouvillian gap with the number of levels. 


\subsection{Semi-classical estimate of steady-state error vs convergence time}
As our protocol has no inherent coherent leakage processes, we expect that the steady state stabilization error will be set primarily by competition between the stabilization process pumping population into the target Bell state $|T\rangle$ and local decoherence leading to decay from the target state. In the absence of decoherence, the system would relax into the target state exponentially at a rate $\tau^{-1}$, so $\dot{\varepsilon}(t)\sim -\tau^{-1}\varepsilon(t)$. If instead the stabilization mechanism were turned off and we consider only the effect of decoherence, at short times the decay out of $|T\rangle$ is exponential with a rate $\gamma\approx T_{B}^{-1}$ leading to $\dot{F}(t)=-\dot{\varepsilon}(t)\sim-T_{B}^{-1}(1 - \varepsilon(t))$. Taking both these processes together, we estimate the steady state error $\varepsilon_\infty$ by solving $\dot{\varepsilon}(t) = 0$, yielding 
    $\varepsilon_\infty \approx \tau/T_{B}$,
when in the dissipation engineering regime where $\tau\ll T_{B}$. As shown in \cite{Supplement}, this estimate accurately predicts the simulated steady state error of the protocol in the absence of detuning- or asymmetry-driven coherent error processes.  
%
\section{Acknowledgments}
%
This material is based upon work supported by the U.S. Department of Energy, Office of Science under award number DE-SC0019461.
\par
%
%
%

\begin{thebibliography}{41}
\bibitem[{eq4()}]{eq4note}
\bibinfo{note}{This assumes that the noise of the channel is independent of the
  initial state (or the channel input) which is a reasonable approximation for
  steady states that are globally asymptotically stable, such as the one
  engineered by the protocol here.}
\end{thebibliography}

%
\section{Author contributions}
T.B., D.R., G.R. and L.R. performed the experiments. T.B. and L.R. analyzed the data and performed simulations with inputs from L.G. and A.K. E.D. developed the protocol and performed theoretical calculations under guidance of A.K. K.C. fabricated the device, J.A. and R.S. developed and performed initial tests on the device. A.K. and L.R. supervised the project. T.B., E.D., A.K., and L.R. wrote the manuscript with inputs from all authors. All authors provided suggestions for the experiment and  data analysis, and contributed to discussion of results.  
\end{document}


\title{Trade off-Free Entanglement Stabilization in a Superconducting Qutrit-Qubit System - \textit{Supplementary Information}}
\author{T.~Brown}
\affiliation{Department of Physics and Applied Physics, University of Massachusetts, Lowell, MA 01854, USA}
\affiliation{Quantum Engineering and Computing, Raytheon BBN Technologies, Cambridge, MA 02138, USA}
\author{E.~Doucet}
\affiliation{Department of Physics and Applied Physics, University of Massachusetts, Lowell, MA 01854, USA}
\author{D.~Rist\`{e}}
\affiliation{Quantum Engineering and Computing, Raytheon BBN Technologies, Cambridge, MA 02138, USA}
\author{G.~Ribeill}
\affiliation{Quantum Engineering and Computing, Raytheon BBN Technologies, Cambridge, MA 02138, USA}
\author{K.~Cicak}
\affiliation{National Institute of Standards and Technology, 325 Broadway, Boulder, CO 80305, USA}
\author{J.~Aumentado}
\affiliation{National Institute of Standards and Technology, 325 Broadway, Boulder, CO 80305, USA}
\author{R.~Simmonds}
\affiliation{National Institute of Standards and Technology, 325 Broadway, Boulder, CO 80305, USA}
\author{L.~Govia}
\affiliation{Quantum Engineering and Computing, Raytheon BBN Technologies, Cambridge, MA 02138, USA}
\author{A.~Kamal}
\affiliation{Department of Physics and Applied Physics, University of Massachusetts, Lowell, MA 01854, USA}
\author{L.~Ranzani*}
\affiliation{Quantum Engineering and Computing, Raytheon BBN Technologies, Cambridge, MA 02138, USA}

\date{\today}
\maketitle
%
%
\section{Circuit Hamiltonian}
%
The circuit is described by the following Hamiltonian,
\begin{eqnarray}
    H &=& \omega_a a^\dagger a + \sum_{k\in \{l,r\}}\left(\omega_k b_k^\dagger b_k + \frac{\alpha_k}{2} b_k^\dagger b_{k} (b_k^\dagger b_k +1) \right) \label{eqn1}\\
    & &  \hspace{28pt} +\,\sum_{k\in \{l,r\}}\hspace{-5pt}g_k(\Phi(t)) (b_k+b_k^\dagger)(a+a^\dagger) + \hspace{-5pt}\sum_{k \in \{l,r\}}\hspace{-5pt}\Omega_k(t)(b_k+b_k^\dagger), \nonumber
\end{eqnarray}
where $\omega_k^{n,n+1} = \omega_{k} + (n+1)\alpha_{k}$ define the transition frequencies of the transmon ${k\in\{l,r\}}$, and $\omega_a$ denotes the center frequency of the resonator. $\Omega_k(t)$ is the amplitude of the direct drives on the transmons. The SQUID mediates a tunable coupling strength, ${g_{k}(\Phi)=L_{sq}(\Phi)\sqrt{\omega_{k}\omega_a}/(2\sqrt{L_{k}L_a})}$, between each transmon and the resonator, where $L_{sq}$ denotes the flux-tunable coupler inductance and $L_k$ is the inductance of the respective transmon~\citep{zakka2011quantum,sirois2015coherent}. We assume that $\Phi(t)$ and $\Omega_k(t)$ are periodic functions of time, so that:
\begin{eqnarray}
g_k(t) &=& \sum_{k\in \{l,r\}}\left(g_k^{01}\cos(\omega_{pk}^{01}t+\phi_k^{01}) + \frac{g_l^{12}}{\sqrt{2}}\cos(\omega_{pl}^{12} t+\phi_l^{12})\right)\\
\Omega_k(t) &=& \sum_{k\in \{l,r\}} \Omega_k^{01}\cos(\omega_{dk}^{01}t+\theta_k).
\end{eqnarray}
We can then eliminate the time dependence in the Hamiltonian by moving to a rotating frame w.r.t. the drives. We write $b_k=\sum_{j}\sqrt{j+1}|j\rangle_k\langle j+1 |$ and perform a unitary transformation described by
\begin{eqnarray}
U(t) = {\rm exp}\left[-i \left(\Delta_{a} a^{\dagger}a + \Delta_{r}^{(1)} |1\rangle_{r}\langle 1|+\Delta_{l}^{(1)} |1\rangle_{l}\langle 1|+ \Delta_{l}^{(2)}|2\rangle_{l}\langle 2| \right)t\right]
\end{eqnarray}
to move into a rotating frame with respect to the free Hamiltonian [first line in Eq.~(\ref{eqn1})]. Keeping only the co-rotating terms for the choice of parametric drive frequencies $\omega_{pk}^{01} = \Delta_{a} - \Delta_{k}^{(1)}$, $\omega_{pl}^{12} = \Delta_{a} - \Delta_{l}^{(2)}$ and Rabi drive frequencies $\omega_{dk}^{01} = \Delta_{k}^{(1)}$, we obtain,
\begin{eqnarray}
H' &=& U^{\dagger}HU - i\hbar U^{\dagger}\frac{dU}{dt} \nonumber\\
&=& \delta_a a^\dagger a + \sum_{k\in \{l,r\}} \delta_k^{01} |1\rangle_k\langle 1| + \delta_{l}^{12} |2\rangle_l\langle 2| + H_{\textrm{I}},
\end{eqnarray}
with
\begin{eqnarray}
\quad H_{\textrm{I}} &=& a^\dagger \left(\frac{g_{l}^{12}}{2} e^{i \phi_{l}^{12}} |1 \rangle_l \langle 2|+\sum_{k\in \{l,r\}} \frac{g_{k}^{01}}{2} e^{i \phi_{k}^{01}} |0 \rangle_k \langle 1| \right) + \sum_{k\in \{l,r\}}\frac{\Omega_k^{01}}{2} e^{i \theta_{k}} |0 \rangle_k \langle 1| + h.c.,
\label{eq:Hdirect}
\end{eqnarray}
Here $\delta_{a}=\Delta_{a}-\omega_a$ is the resonator detuning, $\delta_k^{01}=\Delta_{k}^{(1)}-\omega_k^{01}$ denote the detunings from 0-1 transitions frequencies, and $\delta_{l}^{12}=\Delta_{l}^{(2)}-\omega_{l}^{12}$ is the detuning from 1-2 transition of the qutrit.
\par
In arriving at this Hamiltonian, we chose to consider only the low three levels of the left transmon and the low two levels of the right transmon, giving us an asymmetric qutrit-qubit system. As discussed in the main text and expanded upon in the next section, this is the \emph{minimal} extension to the system Hilbert space which allows exact stabilization of a Bell state. We could instead have chosen to truncate both transmons to the low three levels and work with a qutrit-qutrit system instead. Our scheme generalizes to this case with the addition of another red sideband drive on either the $1$-$2$ and $0$-$2$ transitions of the right transmon to ensure any population in the second excited state decays quickly into the qubit manifold.
%
%
\section{Minimality of qutrit-qubit scheme}
%
In this section we motivate why our scheme is minimal to achieve tradeoff-free stabilization of a two-qubit Bell state. In dissipative stabilization, a quantum system is coupled to its environment so that the desired target state is a stationary state of the dynamics, forming a so called dark state. The conditions for a target state $\ket{\psi}$ to be a dark state are: (i) $\ket{\psi}$ is an eigenstate of the system Hamiltonian, and (ii) $\ket{\psi}$ is a left eigenstate (or null state) of the jump operator, i.e.
\begin{eqnarray}
& & L|\psi\rangle = 0 \quad {\rm and} \quad 
\langle\psi|L \neq 0
\label{eqn:constraints}    
\end{eqnarray}
The first condition in~(\ref{eqn:constraints}) prevents decay \textit{out of} the target state, while the second condition states that the target state is \emph{not} a dark state of the dual (time-reversed) system, \textit{i.e.} there is a decay channel \textit{into} the target state. The latter is necessary to ensure that the desired state can be reached irrespective of the initial conditions of the system. Restricting to only quasi-local jump operators that are realizable with red and blue sideband couplings, the most general form of $\hat{L}$ is,
\begin{eqnarray}
L = \left(c_1^- \sigma_1 + c_2^- \sigma_2 + c_1^+ \sigma_1^\dagger + c_2^+ \sigma_2^\dagger\right),
\end{eqnarray}
where $c_j^\pm$ are arbitrary coefficients set by the amplitudes and phases of sideband drives. 
Satisfying the constraints in Eq.~(\ref{eqn:constraints}), for $\ket{\psi}=|S\rangle = (1/\sqrt{2})(|01\rangle - |10\rangle)$, requires that the sideband drive parameters be chosen such that $c_1^\pm = c_2^\pm$. Unfortunately, this immediately implies that the second constraint in Eq.~(\ref{eqn:constraints}) is impossible to satisfy due to resultant symmetry of the evolution operators as explained in the main text. 
\par
Typical Bell state stabilization protocols employing linear dissipation circumvent this issue by `dressing' the target state. If the target state is not chosen to be the exact singlet state but $(|S\rangle + \delta|\xi\rangle)/\sqrt{1+\delta^2}$ instead, where $\ket{\xi}$ is orthogonal to $|S\rangle$ (e.g. $|gg\rangle$ \cite{Stannigel2012}), then it is possible to find a set of $c_j^\pm$ for which both constraints in Eq.~(\ref{eqn:constraints}) can be satisfied simultaneously. Note that $\delta$ needs to be small, for the singlet fraction in the steady-state to be high, i.e.
\begin{equation}
    F_{ss}(\delta) = \left|\langle S|\psi\rangle\right|^2 = \frac{1}{1+\delta^2} \sim 1-\delta^2
    .
\end{equation}
However, in the limit of $\delta\to 0$ we find that the stabilization rate $\tau^{-1}\to 0$ also. To see this, note that by Fermi's golden rule, the decay rate into $\ket{\psi}$ from $\ket{\phi_j}$ is proportional to the magnitude squared of the matrix element $\langle\psi|L|\phi_j\rangle$. Since $L^{\dagger}|S\rangle = 0$, this must be proportional to $\delta$,
\begin{equation}
    \langle\psi|L|\phi_j\rangle = \frac{\delta}{\sqrt{1+\delta^2}}\langle\xi|L|\phi_j\rangle \sim \delta\langle\xi|L|\phi_j\rangle
    .
\end{equation}
\par
These two results taken together force a tradeoff to be made when operating a protocol based on this approach; thus achieving high fidelity requires that $\delta$ be as small as possible, but this necessarily causes the stabilization rate to drop. In particular the product between error $1-F_{ss}$ and time $\tau$ is a constant. This tradeoff is not an inherent feature of all Bell state stabilization protocols. In a recent work~\citep{Doucet2020}, we theoretically showed that it is possible to stabilize Bell states exactly (up to local decoherence) by employing a strong dispersive coupling between the qubits and the auxiliary cavity to lift the energy degeneracy between the zero and one photon manifold of the qubit-cavity system, thus expanding the set of accessible jump operators. The scheme presented in the main text uses a different approach based on expanding the Hilbert space to include the second excited state of transmons and realize the jump operator shown in Eq.~(2) of the main text.
\begin{figure}[t!]
    \centering
    \includegraphics[width=0.8\textwidth]{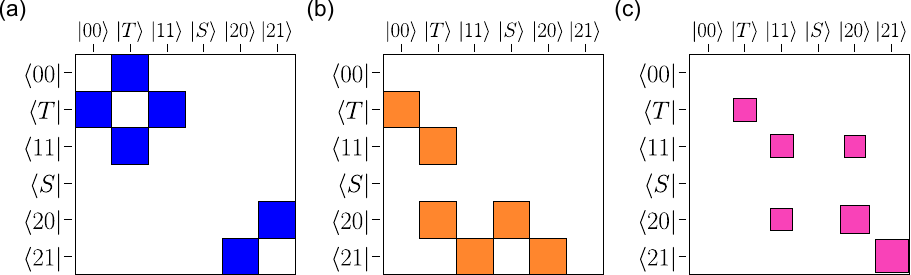}
    \caption{Hinton diagrams of the effective (a) $H$, (b) $L$, and (c) $L^\dagger L$ in the qutrit-qubit space after adiabatic elimination of the resonator.}
    \label{fig:hintons}
\end{figure}
This form of quasi-local dissipation allows the use of unconditional CW interactions to achieve Bell state stabilization, without any tradeoff between fidelity and speed. This is illustrated in Fig.~(2b) of the main text, which shows how the stabilization mechanism of the scheme works by using the extra qutrit levels to allow an indirect path from the triplet manifold into the singlet manifold. The effective Hamiltonian $H$ and jump operator $L$ after adiabatic elimination of the resonator are illustrated in Fig.~\ref{fig:hintons}. Note the off diagonal elements of $L^\dagger L$ which indicate that the transmon-resonator coupling induces a coherent interactions between the two transmons, besides implementing dissipation on different two-qubit states.
%
\section{Extensions to the stabilization scheme}
%
\begin{figure}[b!]
    \centering
    \includegraphics[width=0.5\textwidth]{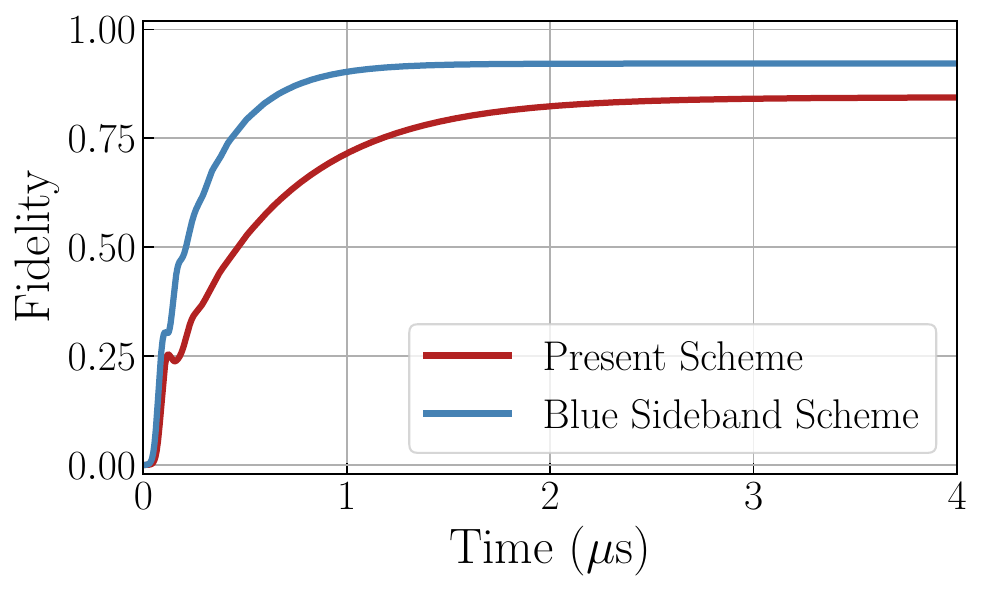}
    \caption{Comparison of stabilization trajectories for current direct drive-based stabilization scheme (red) and a scheme utilizing a two-photon parametric drive on qubits (blue). In both cases, the transmon resonator couplings were set to $g^{01}_{l,r}=2\pi\times7.5\,\si{\mega\hertz}$ and $g^{12}_l=2\pi\times13.1\,\si{\mega\hertz}$. For the present scheme, the optimal direct drive strength used is $\Omega_{l,r}=2\pi\times7.2\,\si{\mega\hertz}$. For the two-photon driving scheme, the optimal coupling strength used is $g_{qq}=2\pi\times15.2\,\si{\mega\hertz}$.}
    \label{fig:rsbbsb}
\end{figure}
While not used in the present work, our device design allows parametric transmon-transmon interactions in addition to the parametric transmon-resonator interactions employed for dissipation engineering. In fact it is in principle possible to design a purely parametric stabilization protocol according to the following Hamiltonian:
\begin{eqnarray}
    H &=& \omega_a a^\dagger a + \sum_{k\in \{l,r\}}\left(\omega_k b_k^\dagger b_k + \frac{\alpha_k}{2} b_k^\dagger b_{k} (b_k^\dagger b_k +1) \right) \\
    & &  \hspace{27pt} +\,\sum_{k\in \{l,r\}}\hspace{-5pt}g_k(\Phi(t)) (b_k+b_k^\dagger)(a+a^\dagger) + g_{qq}(\Phi(t))(b_l + b_l^\dagger)(b_r + b_r^\dagger)
    ,
\end{eqnarray}
where additional flux drives are included such that $g_{qq}(t)$ can drive transmon-transmon interactions on resonance. In particular our scheme can be made purely parametric by use of a transmon-transmon blue sideband drive at $\omega_{lr}^+ = \omega_{l}^{01} + \omega_{r}^{01}$ in lieu of the direct transmon drives in the present scheme, yielding a purely parametric stabilization scheme with several advantages over the version built around direct driving. The effective interaction Hamiltonian for such a scheme is,
\begin{eqnarray}
\quad H_{\textrm{I}} &=& a^\dagger \left(\frac{g_{l}^{12}}{2} e^{i \phi_{l}^{12}} |1 \rangle_l \langle 2|+\sum_{k\in \{l,r\}} \frac{g_{k}^{01}}{2} e^{i \phi_{k}^{01}} |0 \rangle_k \langle 1| \right) + \frac{g_{qq}}{2} |00 \rangle \langle 11| + h.c.
\label{eq:Htwophoton}
\end{eqnarray}
Using exactly the same sideband coupling strengths and an optimized transmon-transmon coupling, we would expect to achieve a fidelity of 92\%, as compared to 84\% achieved using direct drives, on the same device as illustrated by the simulations presented in Fig.~\ref{fig:rsbbsb}. Such two-photon interaction was not available in our device, because stray capacitive coupling between the flux line and the SQUID limited the available bandwidth, but it should be accessible with relatively straightforward design modifications. Further optimization of transmon-resonator coupling strengths over a wide range of parameters (including the case when the transmon-transmon drive is comparable in strength to the direct drive strengths currently used), we expect a $2\times$-$3\times$ improvement in both stabilization time and error from the inclusion of parametric qubit-qubit interactions.
%
%
\section{Experimental Setup and Characterization}
%
\begin{figure}[b!]
    \centering
    \includegraphics[width=0.8\textwidth]{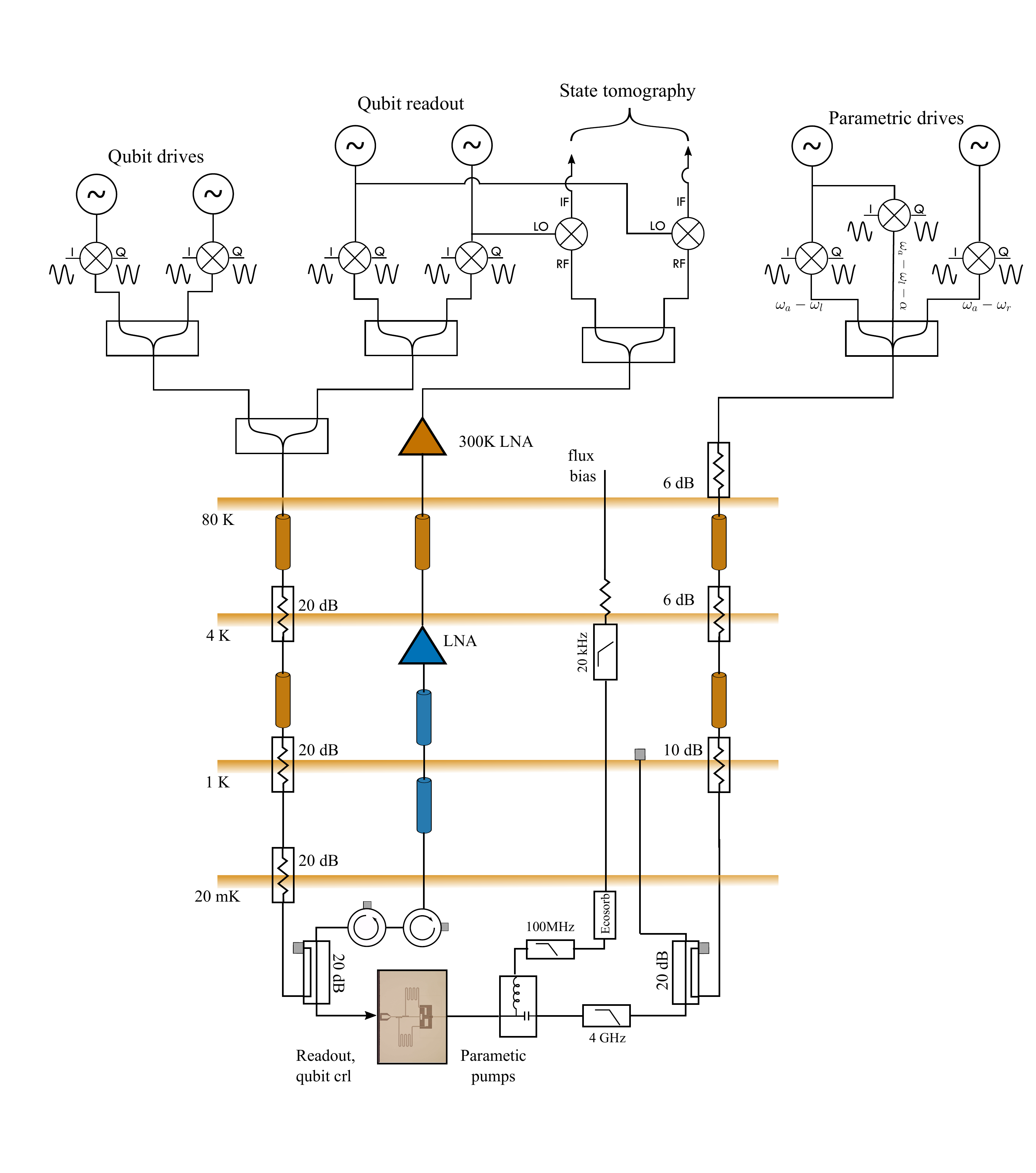}
    \caption{Experimental Setup}
    \label{fig:ExpSetup}
\end{figure}
The experimental setup is shown in Fig.~\ref{fig:ExpSetup}. The sample is enclosed in an Aluminum box, wirebonded to coaxial connectors and mounted at the 20mK stage of a dilution refrigerator. Qubit readout and control signals are injected via a single heavily attenuated microwave line and the reflected qubit readout signals are amplified, demodulated and integrated with a matched filter~\citep{ryan2017hardware}. The parametric drives are combined, attenuated by 42dB and lowpass filtered before reaching the SQUID flux port. The \SI{4}{\GHz} lowpass filter at base temperature protects the qubits from noise and decay through the pump line. We further minimized the length of coaxial cables between the device and filter stages to avoid standing wave resonances near the qubit frequencies. The qubit device is further enclosed into a $\mu$-metal shield to protect it from stray magnetic fields.
\par
The SQUID flux bias signal is lowpass filtered via a 3-stage RC network at 4~K and an Ecosorb filter at base temperature.   We used a multi-channel signal generator for the control and parametric drives to guarantee relative phase stability. The microwave tones are then sideband modulated by fast arbitrary waveform generators~\citep{ryan2017hardware} to obtain the desired signal frequencies.  
%
%
\subsection{Device Characterization}
%
We characterized the 0-1 and 1-2 transition frequencies for transmons using pulsed microwave spectroscopy [Fig.~1 of the main text], and extracted Josephson energy $E_J$ and charging energy $E_c$~\citep{koch2007charge} of the junctions using standard formulae. Furthermore, we measured the relaxation and dephasing rates of the three lowest transmon levels by performing inversion recovery and Ramsey experiments~\citep{peterer2015decay}. For the $1-2$ transition we verified that the $\gamma^{12} \approx 2\gamma_{01}$ in accordance with the dipole matrix element for a transmon-type oscillator, by fitting the measured decay curves for the three levels to a sequential decay process $2\rightarrow 1\rightarrow 0$. To extract the dephasing time for $0-2$ and $1-2$ transitions, we first subtracted exponential background decay to level $|1\rangle_{k}$ and then fitted the measured oscillations to a decaying sinusoid. The device parameters are summarized in Table~\ref{tab:qbtparameters}.
\begin{table}[h!]
\begin{center}
\bgroup
\def\arraystretch{1.2}
\begin{tabular}[t]{p{6cm}p{9cm}}
\noalign{\hrule height 1pt}
Parameter & {\bf Measured Value} \\
\hline
\hline
Josephson Energy ($E_{J}/2\pi$) & $l$ = 24.4 GHz, \hspace{5pt} $r$ = 20.9 GHz, \hspace{5pt} ${\rm SQUID}$  $\approx$ 340 GHz\\
Anharmonicity ($\alpha_{l,r}/2\pi$) &  $l=  -198$ MHz, \hspace{5pt}
$r= -164$ MHz\\
Rabi drive amplitude ($\Omega_{l,r}^{01}/2\pi$) & 7.2 MHz \\
0-1 sideband amplitude ($g_{l,r}^{01}/2\pi$) & 7.5 MHz \\
1-2 sideband amplitude ($g_{l}^{12}/2\pi$) & 13.1 MHz\\
Resonator decay rate ($\kappa/2\pi$) & 4.73 MHz\\
0-1 relaxation rate ($\gamma^{01}/2\pi$) & $l=33.9$ \si{\kilo\Hz}, \hspace{5pt} $r=27.5$ \si{\kilo\Hz} \\
1-2 relaxation rate ($\gamma^{12}/2\pi$) & $l=69.2$ \si{\kilo\Hz}  \hspace{5pt} $r=56.9$ \si{\kilo\Hz}
\\
0-1 dephasing rate ($\gamma^{11}/2\pi$) & $l=18.8$ \si{\kilo\Hz}, \hspace{5pt} $r=13.9$ \si{\kilo\Hz} \\
0-2 dephasing rate ($\gamma^{22}/2\pi$) & $l=43$ \si{\kilo\Hz} \hspace{5pt} $r=32.8$ \si{\kilo\Hz}
\\
\noalign{\hrule height 1pt}
\end{tabular}%
\caption{Summary of measured device parameters.}
\label{tab:qbtparameters}
\egroup
\end{center}
\end{table}
\begin{figure}[h!]
    \centering
    \includegraphics[width=246pt]{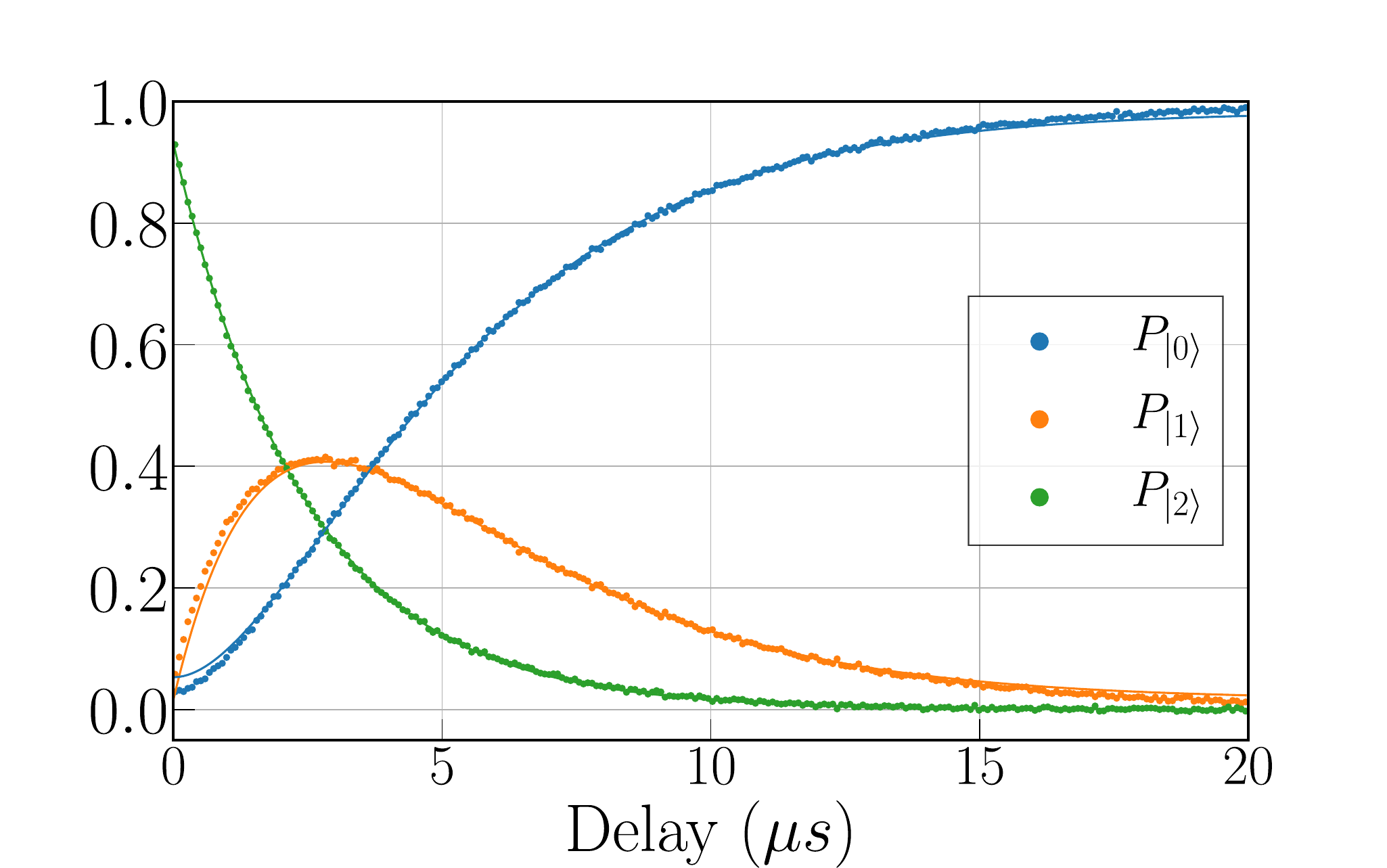}
    \includegraphics[width=246pt]{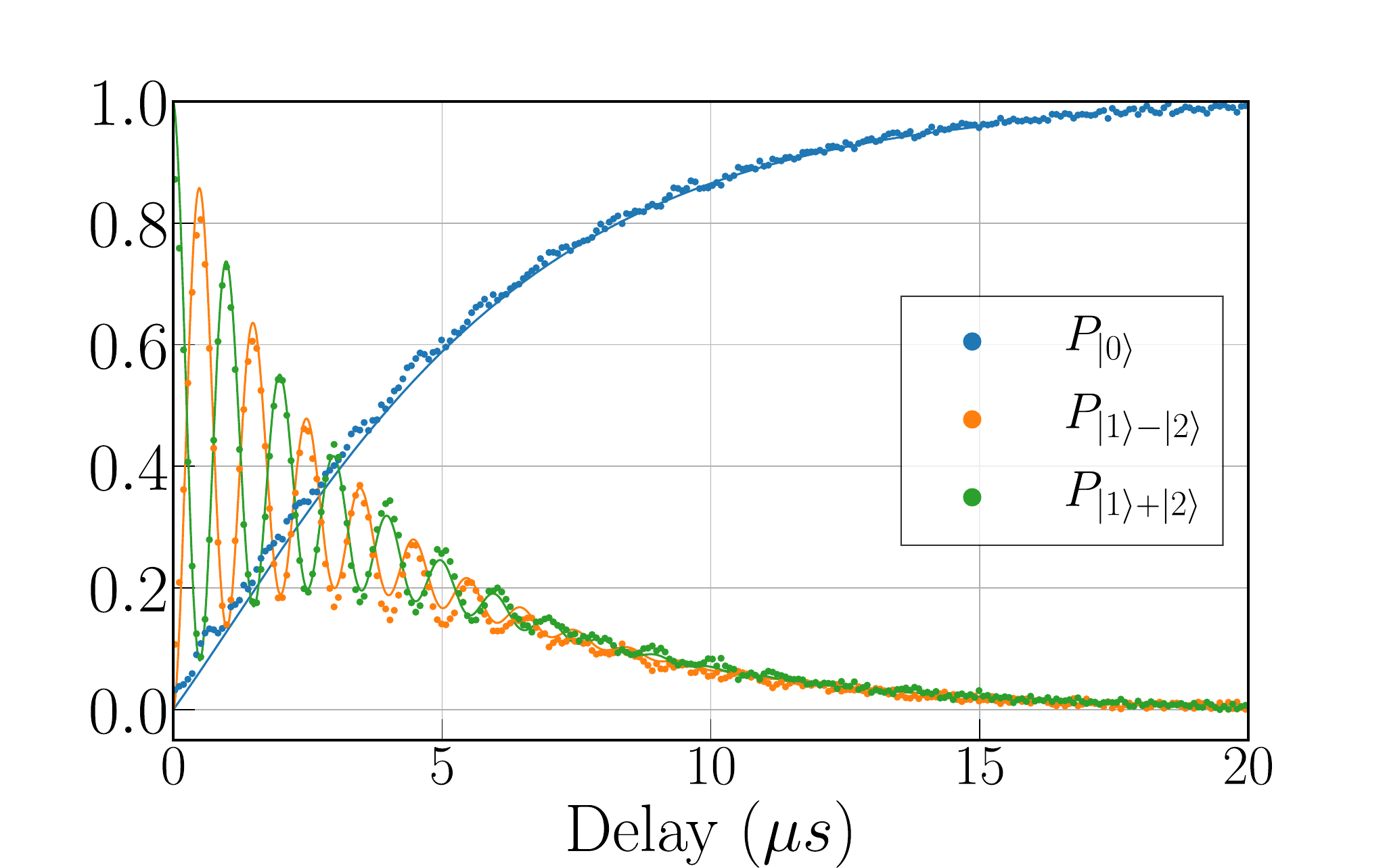}
    \caption{Left: Populations in $|0\rangle$, $|1\rangle$ and $|2\rangle$ in a transmon qutrit prepared in $|2\rangle$ and allowed to relax. Population in $|0\rangle$ deviates from the data due to measurement errors leading to a slightly reduced trace. Right: $|0\rangle$, $(|1\rangle-|2\rangle)/\sqrt{2}$ and $(|1\rangle+|2\rangle)/\sqrt{2}$ populations tracked over time in a Ramsey experiment with the transmon qutrit initialized in $(|1\rangle + |2\rangle)/\sqrt{2}$. In both plots, experimental data is shown as dots and master equation simulations are shown as lines.}
    \label{fig:fdecay}
\end{figure}
\par
The envelope of the Ramsey oscillations is exponential indicating that dephasing is dominated by white noise. Under the assumption that this white noise component was due entirely to photon shot noise in the center resonator, we can infer the (residual) photon population $\langle n \rangle$ in the resonator in the weak dispersive regime (since $\chi/\kappa < 1$); using $\gamma^{11} = 8\langle n \rangle \chi^{2}/(\kappa + 4\chi^{2}/\kappa)$ \cite{Gambetta2006}, we obtain an average photon number $\langle n \rangle \approx 0.02$ consistent with thermal populations reported previously in other cQED setups. In the absence of correlated noise (for instance, $1/f$-type), the dephasing rate of the 1-2 transition can be approximated as $\gamma_{\phi}^{12} \approx \gamma^{22} + \gamma^{11}$ \cite{li2011decoherence}; this describes the dephasing of 1-2 Ramsey oscillations shown in Fig.~\ref{fig:fdecay}. 
%
%
\subsection{Stabilization Drives Tuning Procedure}
%
\begin{figure}[b!]
    \centering
    \includegraphics[width=0.95\textwidth]{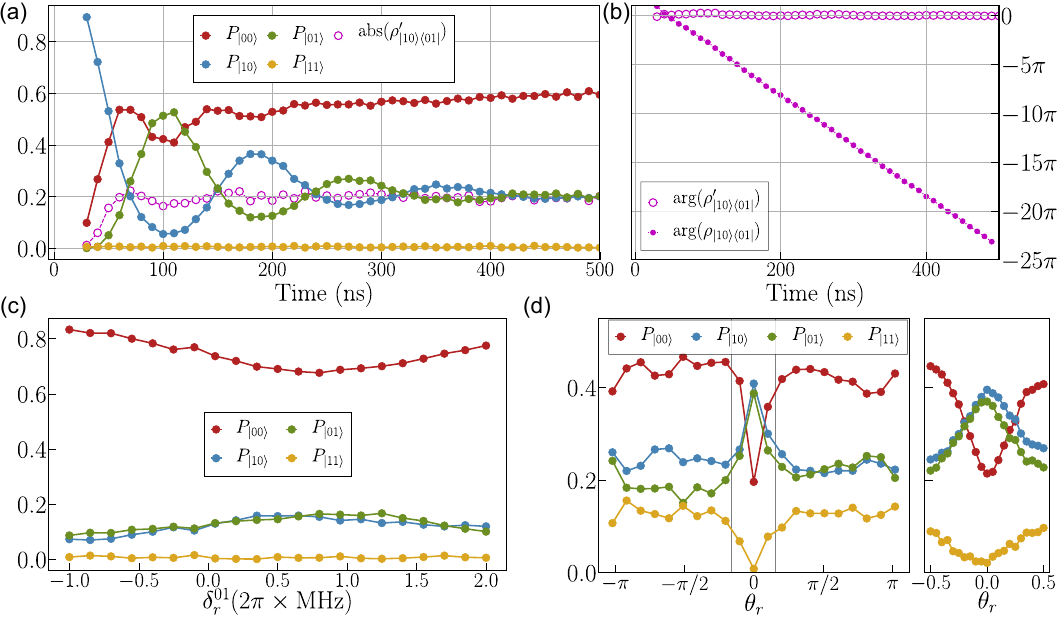}
    \caption{(a) Evolution of the state populations when initializing in $|10\rangle$ under just parametric drives. $P_{|00\rangle}$ is minimized when the drives are on resonance. (b) AC Stark shift of the qubits introduces a Z-rotation of the state $\rho$. Such rotation is compensated by adjusting the drive phases, resulting in the state $\rho'$. (c) Effect of parametric drive detuning on $P_{|00\rangle}$. (d) When the Rabi drives are introduced the Rabi and parametric drive phases need to be aligned to stabilize the Bell state. When the phases are aligned the qubit population in the even-parity manifold is minimized.}
    \label{fig:Tuning}
\end{figure}
%
Here we describe our procedure to tune the stabilization drives together to compensate for pump-induced Stark shifts of the qubit and resonator frequencies due to the SQUID nonlinear inductance~\cite{lu2017universal}. Most tuning steps require only knowledge of the diagonal part of the density matrix, which we extracted directly from expectation values of $\langle \sigma_{zl,r} \rangle$ and the correlator $\langle \sigma_{zl}\sigma_{zr}\rangle$ (see section on Quantum State Tomography later).
\par
We initially calibrate each drive independently as explained in the main text and use the resulting frequencies and amplitudes as our starting point. We then initialize the system in either $|01\rangle$ or $|10\rangle$ and apply the parametric drives for a time $t\gg 1/\kappa$, but much lower than the qubit decoherence times, see Fig.~\ref{fig:Tuning}(a). We verified from our simulations that at such time the population in the even parity manifold reaches a minimum when the drives are on resonance, allowing us to determine each drive frequency [Fig.~\ref{fig:Tuning}(c)]. Moreover, the ratio $P_{10}/P_{01}$ between the qubit state populations quickly converges toward the ratio $(g_r^{01}/g_l^{01})^{2}$ consistent with the steady state in the presence of asymmetric 0-1 sideband amplitudes,
\begin{equation}
    \rho_{ss} = \alpha |00\rangle\langle 00| + \beta |\psi_{\rm odd}\rangle\langle \psi_{\rm odd}|,
\end{equation}
where $|\psi_{\rm odd}\rangle = (g_{r}^{01}|10\rangle - g_{l}^{01}|01\rangle)/\sqrt{(g_{l}^{01})^{2} + (g_{r}^{01})^{2}}$.
This allows fine tuning of 0-1 sideband amplitudes by minimizing the difference between $P_{10}, P_{01}$ in the steady state. Note that while the steady state of the system in the presence of all stabilization drives is independent of the initial state of the qubits, this is not the case in the absence of Rabi drives.
\par 
At this point of the tuning process the third parametric drive at $\omega_a - \omega_{l}^{12}$ affects the qubit and resonator frequencies but does not produce any transition, because of photon number conservation. In order to tune the third parametric drive, we initialize the qubits in $|20\rangle$, so that the $|20\rangle|0\rangle_a$ to $|10\rangle|1\rangle_a$ swap process can be observed. We subsequently minimize the expectation value of $\langle \sigma_{zr} \rangle$ as a function of the drive frequency at $t\approx \SI{150}{\nano\second}$, which corresponds to a population transfer between the transmons with maximum efficiency. Once all parametric drives are tuned on resonance, we perform full state tomography of the qubits over time and measure the state $Z$-rotation caused by the pump-induced AC Stark shift, see Fig.~\ref{fig:Tuning}(b). Such $Z$-rotation is computed accurately from the time evolution of the phase of the $\langle10|\rho(t)|01\rangle$ element of the density matrix and it is equal to $\delta_st$, where $\delta_s=\delta_l^{01}+\delta_r^{01}$ is the sum of the qubit frequency shifts. 
\par 
We finally proceed to tune the Rabi drives. From the previous steps we know the Rabi drive common-mode frequency offset $\delta_s$ as well as their differential offset, since the latter has to match the relative frequency offset of the $g_{l,r}^{01}$ parametric drives. We then proceed to sweep the Rabi drives relative phase and amplitudes until the population of the desired Bell state $|T\rangle$ reaches a maximum, see Fig.~\ref{fig:Tuning}(d). Finally, we further compensate the state $Z-$rotation by adjusting the phase of the parametric pumps and Rabi drives as a function of the pulse width. For long stabilization times we observe a small residual detuning of the order of \SI{50}{\kHz}, which we correct in post-processing [Fig.~\ref{fig:Tuning}(b)]. 
%
%
\subsection{Quantum State Tomography}

We used Quantum State Tomography (QST) to characterize the stabilized state~\citep{ryan2015tomography}. We repeatedly prepared the stabilized state by turning on the stabilization drives for a fixed amount of time and then performed a set of single qubit rotations followed by simultaneous readout of both qubits with separate cavities, with readout fidelity of 86\%. The tomographic gate set consisted of idle, $X_{\pi}$, $X_{\pm \pi/2}$ and $Y_{\pm \pi/2}$ gates on each qubit, for a total of 36 gate combinations. While the minimal gate set for two-qubit QST consists of 16 gate combinations~\citep{filipp2009two,steffen2006measurement}, 36 gate combinations provide an overdetermined set of equations that reduce reconstruction errors. We then measured the expectation values of single-qubit polarization $\langle Z_{k}\rangle = |0\rangle_{k}\langle 0| - |1\rangle_{k}\langle 1|$, as well as the product $\langle Z_{l}Z_{r} \rangle$ by correlating the single measurement shots from each qubit and averaging. For each tomographic measurement we averaged over $8\times10^4$ records. We used a least squares optimizer, weighed by the variance of the observations, to reconstruct the density matrix. The optimizer minimizes the quantity~\citep{ryan2015tomography}
\begin{equation}
    E(\rho) = \sum{(M_j-P_j(\rho))^2/var(M_j)},
\end{equation}
where $M_j$ are the measured single and two-body observables and $P_j$ are the predicted values for a two-qubit state defined by he density matrix $\rho$. Weighing by the variance of the data reduces bias in the reconstructed matrix elements by taking into account that the errors in the measured single- and two-body terms have different standard deviation.
\par
\textit{Approximate calculation of qubit populations} - To tune up the drives we compute only the diagonal elements of the density matrix, without performing a full tomographic reconstruction. The diagonal elements are in fact directly affected by both amplitude imbalance or frequency detuning of the drives, as detailed in the previous section on tuning. We can compute the diagonal elements of $\rho$ directly by measuring $\langle Z_{l}\rangle$, $\langle Z_{r} \rangle$ and $\langle Z_{l}Z_{r} \rangle$ and directly computed the population of the four basis states $|00\rangle$, $|01\rangle$, $|10\rangle$, $|11\rangle$ as,
\begin{align*}
   P_{00} &= \frac{I+\langle Z_{l}\rangle+\langle Z_{r}\rangle+\langle Z_{l}Z_{r}\rangle}{4}, \\
   P_{01} &= \frac{I+\langle Z_{l}\rangle-\langle Z_{r}\rangle-\langle Z_{l}Z_{r}\rangle}{4},\\
   P_{10} &= \frac{I-\langle Z_{l}\rangle+\langle Z_{r}\rangle-\langle Z_{l}Z_{r}\rangle}{4}, \\
   P_{11} &= \frac{I-\langle Z_{l}\rangle-\langle Z_{r}\rangle+\langle Z_{l}Z_{r}\rangle}{4}.
\end{align*}
Note that because of the crosstalk between the two measurement channels, the measurement records are in general a linear combination of $Z$-measurement operators. We calibrated and removed cross-talk between the readout signals by measuring the readout levels for all four basis states and performing a linear inversion.
%
\subsection{Stabilization from Different Initial States}
\begin{figure}[h!]
    \centering
    \includegraphics[width=246pt]{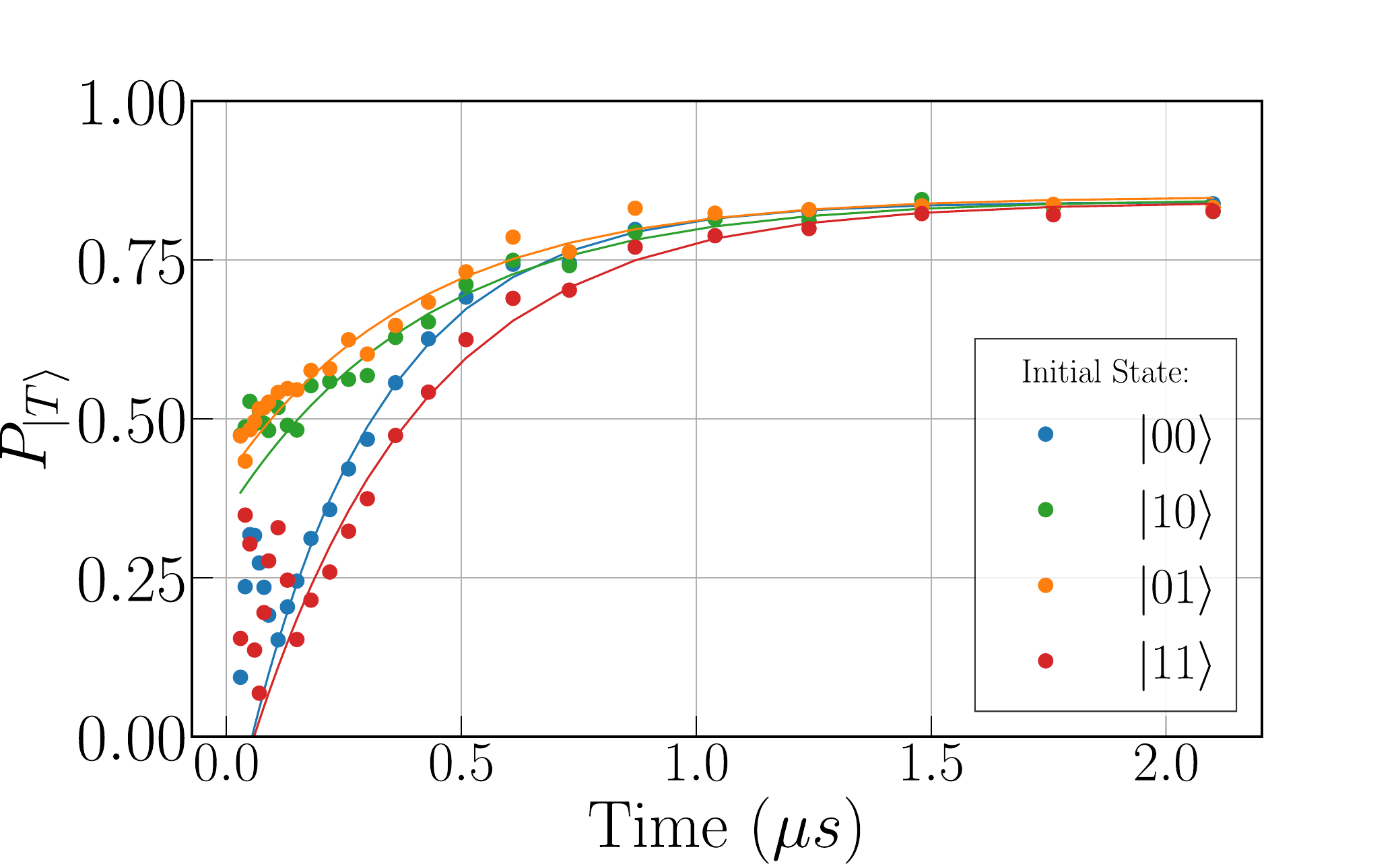}
\caption{Stabilization trajectory of the qutrit-qubit system into the Bell state $|T\rangle$ with the two transmons initialized into each of four states: $|00\rangle, |10\rangle, |01\rangle$ and $|11\rangle$. Measured data is presented as dots, with fits as solid lines. This demonstrates the universality of this stabilization protocol for states in the qubit-qubit space.}
    \label{fig:initstates}
\end{figure}

%
One of the features of our dissipative stabilization protocol is that it is initial state-independent by construction. Any initial state will relax to the engineered steady state of the resulting Liouvillian, allowing for universal stabilization. This is possible because every state in the two-qubit Hilbert space is coupled to the target state via a combination of fast coherent and dissipative channels. This is demonstrated for the stabilization scheme implemented in the main text in Fig.~\ref{fig:initstates}. Note that when the qubits are initialzied in any of the four basis states, the system converges to the same target state with similar fidelity. Moreover an important feature of our protocol is that the convergence time is in fact independent from the initial state.
%
%
%
%
\section{Experimental Imperfections and Crosstalk}
%
\subsection{Robustness to parameter deviations}
%
\begin{figure}[b!]
    \centering
    \includegraphics[width=0.8\textwidth]{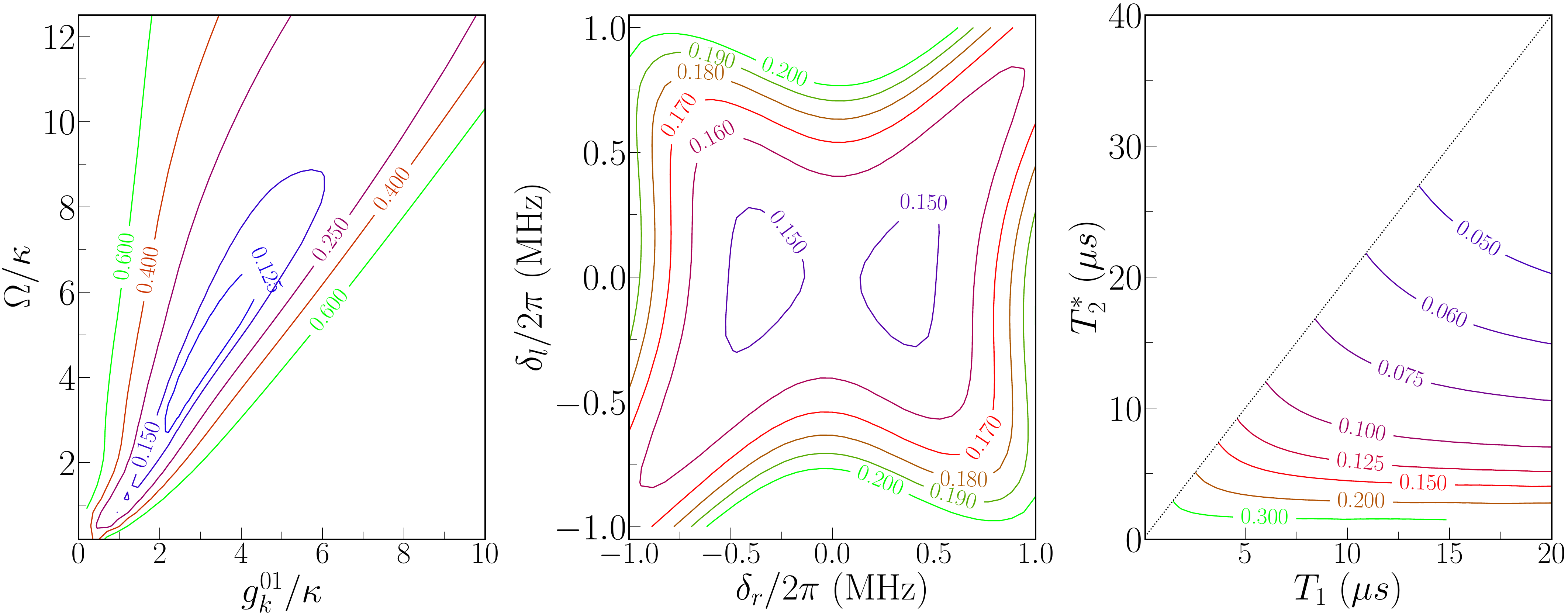}
    \caption{Steady-state error as a function of (a) 0-1 direct drive amplitude and 0-1 sideband drive amplitude, both normalized to resonator linewidth $\kappa$, (b) detunings on the 0-1 transitions for left and right transmons, and (c) transverse and longitudinal relaxation times (assuming same decoherence rates for both transmons). The other parameters were fixed at the nominal values reported in Table S1.}
    \label{fig:contourplots}
\end{figure}
We simulated the sensitivity of our stabilization protocol to parameter imperfections and show the results in Fig.~\ref{fig:contourplots}. The protocol is relatively robust to drive detuning, so that a detuning of up to \SI{1}{\MHz} causes a drop in target state fidelity of only $4\%$. By improving the qubit coherence up to $T_1=T_2^*\approx 20\mu s$, fidelities above 95\% are achievable in the current scheme. 
%
\subsection{Impact of leakage to higher transmon levels}
%
%
\begin{figure}[b!]
    \centering
    \includegraphics[width=0.95\textwidth]{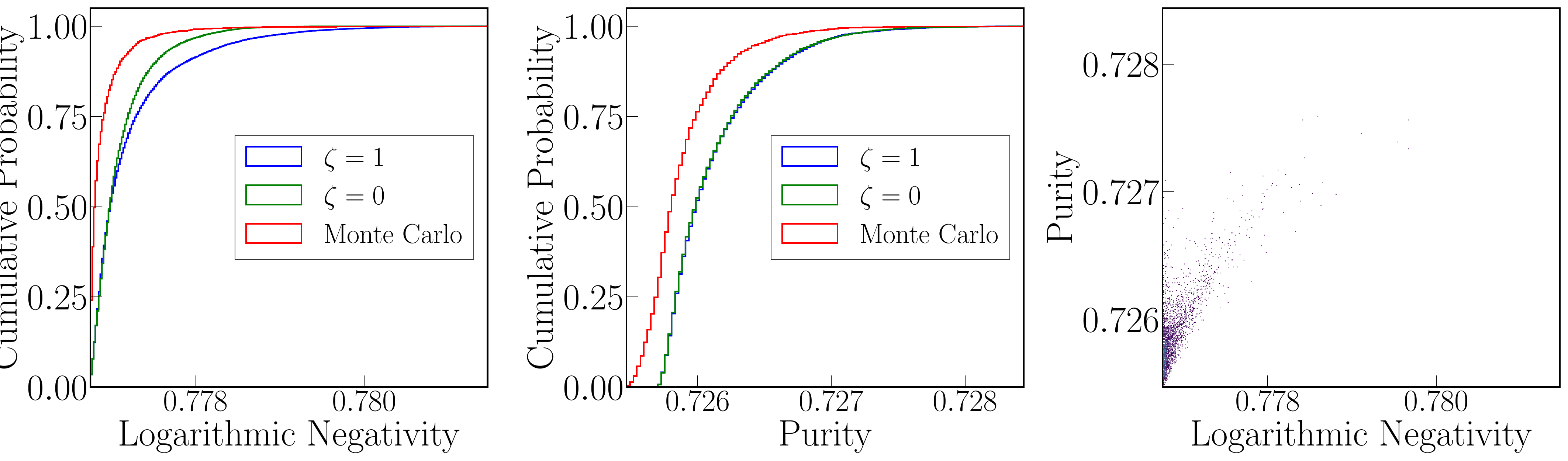}
    \caption{Results of the Monte-Carlo entanglement bound study. Left, Center: logarithmic negativity and purity respectively for cases where the higher level population balance parameter $\zeta$ is randomized $\zeta$ case, as well as for the cases where $\zeta=0$ and $\zeta=1$. Right: histogram of the purity and logarithmic negativity based on Monte Carlo extension of the density operator.}
    \label{fig:tanglebound}
\end{figure}
In the main text we characterized our system by performing two-qubit quantum state tomography. This approach can be considered valid if we have negligible leakage out of the computational space into higher transmon levels. In particular leakage into the $|20\rangle$ and $|21\rangle$ energy levels is expected to be the dominant mechanism, because of the presence of the sideband drive targeting the 1-2 transition of qubit $l$. In order to determine the impact of such leakage, we measured the population of the second excited state $|2\rangle_l$ during stabilization by following the procedure in~\citep{peterer2015decay}. We can then use this information to constrain the trace of the tomographic reconstruction and estimate error bounds caused by leakage. Note that the latter measurement does not contain information about the $|20\rangle$ and $|21\rangle$ level populations independently, which would require a full 3-level tomographic reconstruction. We investigate the possible impact of leakage as well as potential unknown correlations in the qubit-qubitrit Hilbert by use of a Monte Carlo approach to examine the distribution of the log-negativity over the space of $6\times6$ qubit-qutrit density matrices that are compatible with the reconstructed qubit-qubit density matrix and measured leakage to $|2\rangle_{l}$, as explained below.
\par
In the Monte Carlo study we first randomly choose the leakage population between the two levels, as shown in Eq.~(\ref{eqn:fbal}) with random real variable $\zeta \in [0,1]$,  
\begin{equation}
    \tilde{\rho} = \rho_{qq} + (1-\textrm{Tr}(\rho)) ( (1-\zeta)|20\rangle \langle 20| + \zeta|21\rangle \langle 21| ),
    \label{eqn:fbal}
\end{equation}
where $\rho_{qq}$ is the reconstructed density matrix from two-qubit state tomography. Eq.~(\ref{eqn:fbal}) assumes that there are no residual correlations outside the qubit manifold, which we expect to be true to a good approximation as we approach the target state. Possible remaining correlations between the population in the qutrit $|2\rangle$ level may have an impact on the estimate of the state entanglement. To this end, we modify the density matrix in Eq.~(\ref{eqn:fbal}) by adding correlations between the qubit and qutrit sectors, weighted with a randomized complex variable $C_{(\Phi,\Psi)}$ with $|C_{(\Phi,\Psi)}| \in [0,1)$ and $\arg(C_{(\Phi,\Psi)})\in[0,2\pi)$,  
\begin{equation}
    \tilde{\rho}' = \tilde{\rho} + \sum_{\substack{\Phi \in S_{qq}, \\ \Psi \in S_{t}}} \sqrt{\Phi^\dagger \tilde{\rho}\Phi \times \Psi^\dagger\tilde{\rho}\Psi} \big(C_{(\Phi,\Psi)}\Phi \Psi^\dagger + C_{(\Phi,\Psi)}^\dagger \Psi \Phi^\dagger\big) 
    \label{eqn:tanglebound}
\end{equation}
where $S_{qq} = \{|00\rangle, |01\rangle, |10\rangle, |11\rangle\}$ and  $S_{t} = \{|20\rangle, |21\rangle\}$. The validity of the resulting density matrix is checked by verifying that it is positive semi-definite within a margin of error of $1\times10^{-3}$. We apply this Monte Carlo analysis to the stabilized state density matrix measured at $t=$\SI{50}{\micro\second} [shown in the main text] using the method described above and compute the resulting logarithmic negativity and purity. The results of this analysis are shown in Fig.~\ref{fig:tanglebound}. Our analysis confirms that the qutrit-qubit entanglement is extremely weak as most of the weight of the $6\times6$ density operators clusters near the value of logarithmic negativity of 77\% consistent with the values reported in the main text based on the $4\times4$ qubit-qubit reconstruction. The change in purity is entirely accounted by leakage-dependent reduction of the trace by 2\% (leading to a concomitant reduction in purity of $\rho_{qq}$). Furthermore, the maximum bound of entanglement is within 1\% of the reported value as shown by the span of logarithmic negativity for the entire distribution of constructed density operators.
%
%
%
\subsection{Impact of counter-rotating terms}
\begin{figure}[h!]
    \centering
    \includegraphics[width=0.5\textwidth]{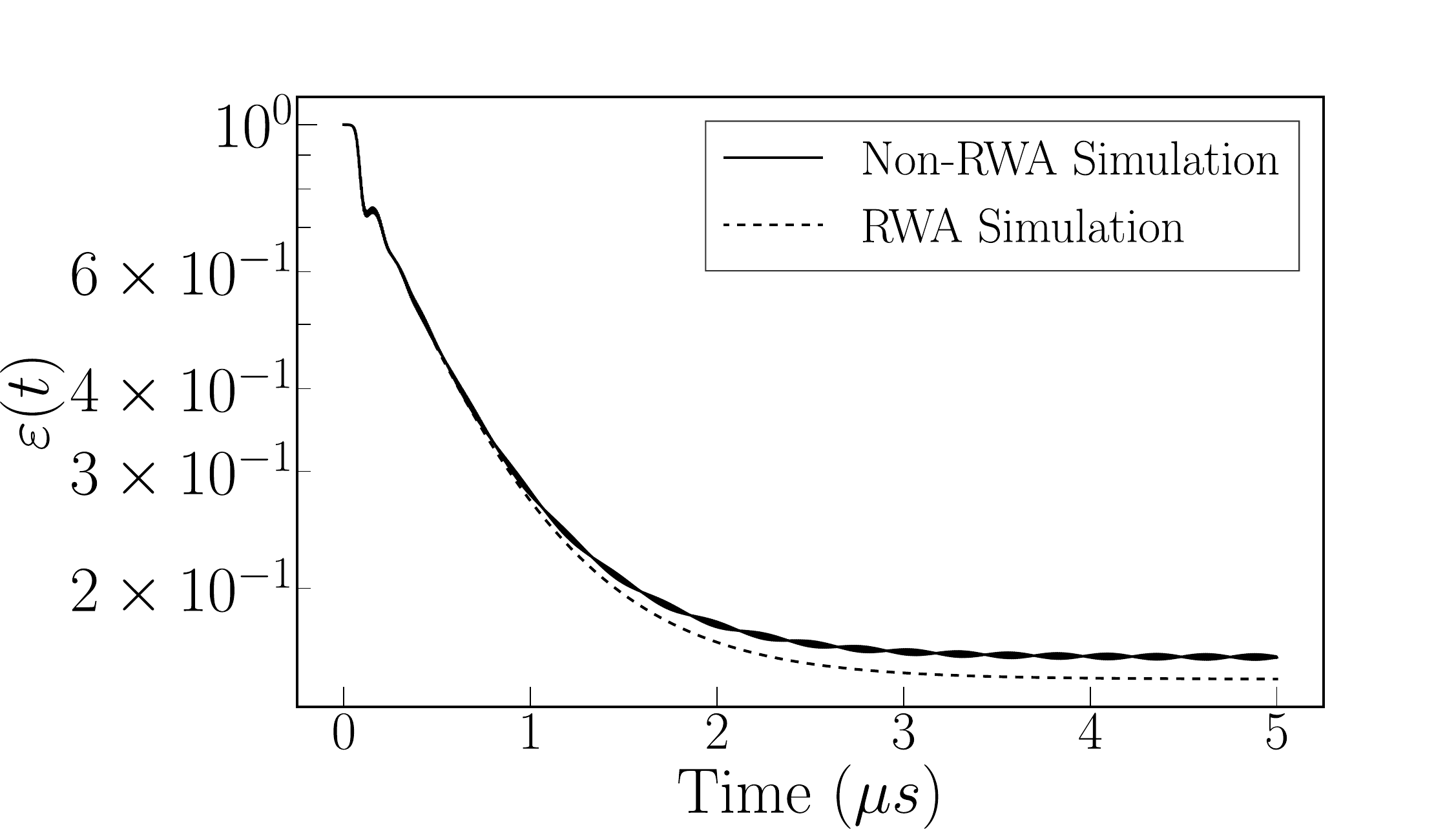}
\caption{Comparison of stabilization performance using RWA Hamiltonian (Eq.~(1) in the main text) and simulations including Eq.~(\ref{eqn:HCR}). The average steady-state fidelity changes from 85.2\% to 84.0\% upon including counter-rotating terms, a small change which is within the margin of measurement and reconstruction error.} 
    \label{fig:CRfig}
\end{figure}

%
Off-resonant terms beyond the rotating-wave approximation have a negative impact on the performance of the stabilization scheme, by introducing coherent leakage out of the target state. The dominant counter-rotating contributions in our stabilization scheme are described by adding the following terms to the interaction Hamiltonian
\begin{equation}
    \begin{split}
H_{\rm CR} &= H_I + a^\dagger\bigg(\frac{g_l^{12}}{2\sqrt{2}} e^{i\phi_l^{12}} e^{-i\alpha_l t}|0\rangle_l\langle 1| + \sum_{k=l,r} \frac{g_k^{01}}{\sqrt{2}}e^{i\phi_k^{01}}e^{+i\alpha_k t}|1\rangle_k \langle 2|\bigg)\\
            & \hspace{25pt} + \sum_{k\in {l,r}} \frac{\Omega^{01}_k}{\sqrt{2}} e^{i\theta_k} e^{+i\alpha_k t}|1\rangle_k \langle 2| +h.c.
\label{eqn:HCR}
\end{split}
\end{equation}
As it is evident in the above formula, the induced leakage is off-resonant by the anharmonicities $\alpha_{l,r}$ of the transmons, which is at least an order of magnitude larger than typical $\chi$-induced leakage in protocols using number-selective \cite{Doucet2020} or state-selective driving \cite{Leghtas2013}. Accordingly, as shown in Fig.~\ref{fig:CRfig}, the effect of these terms is strongly suppressed even for parametric drive amplitudes in excess of 10 MHz.
%
%
\subsection{Effect of crosstalk on error-time scaling}
%
%
\begin{figure}[h!]
\begin{subfigure}[]{\includegraphics[width=0.44\textwidth] {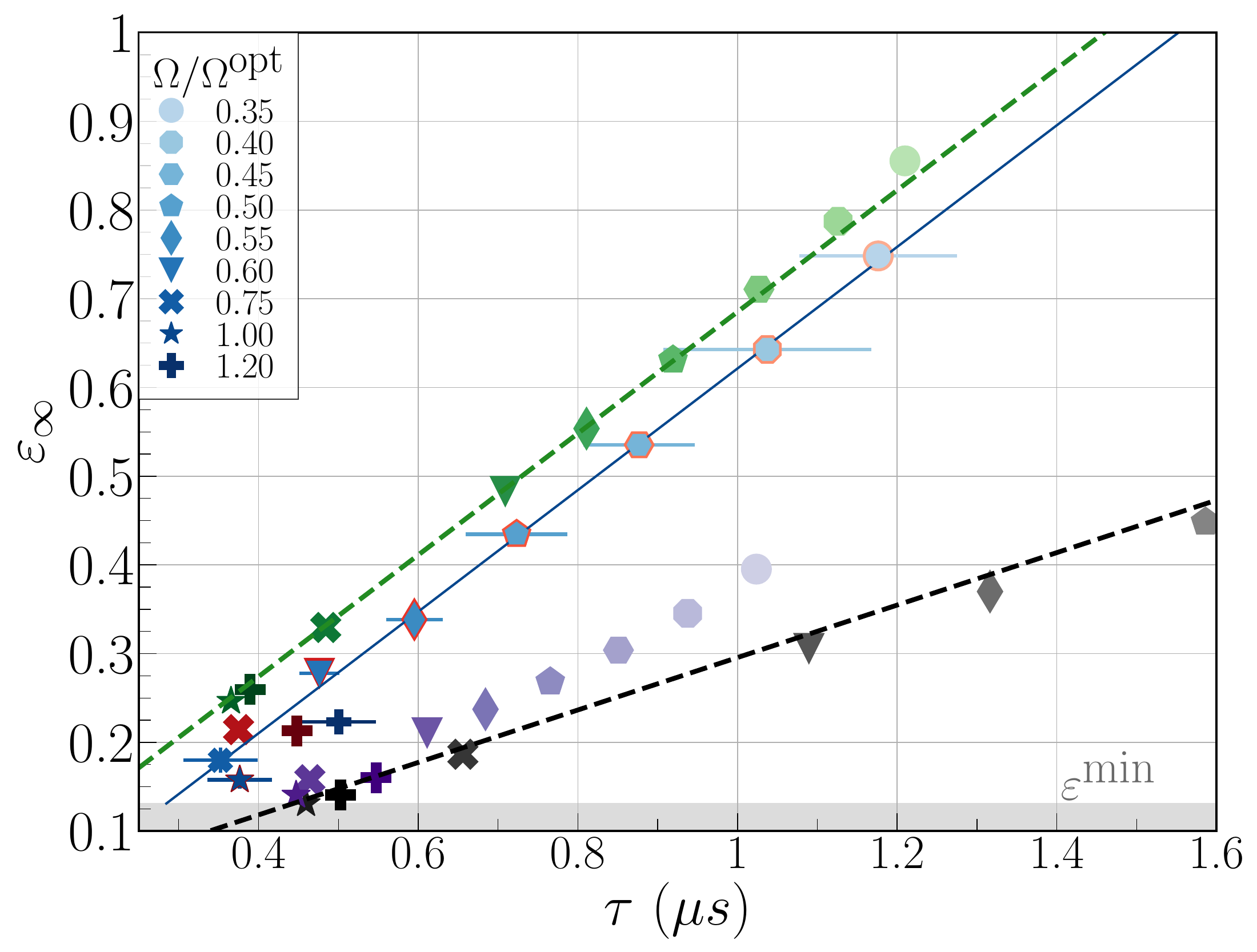}
\label{fig:subfig1}
}%
\end{subfigure}%
\begin{subfigure}[]{\includegraphics[width=0.54\textwidth]{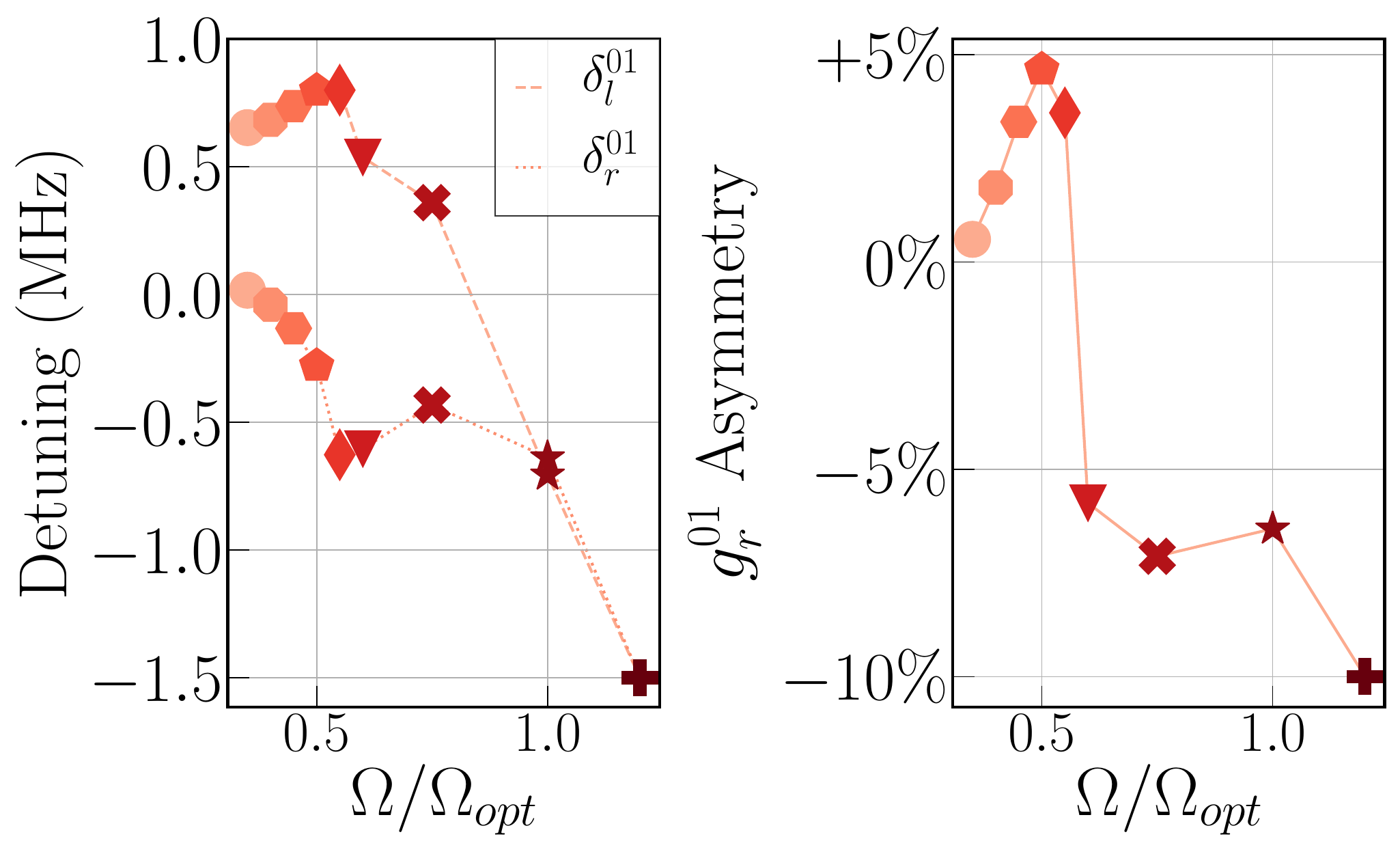}
\label{fig:subfig2}
}%
\end{subfigure}%
\caption{(a) Comparison of data shown in Fig.~4(b) of the main text, with simulations of the full qutrit-qubit and resonator system. Black markers correspond to master equation simulation shown in the Methods, using the experimental parameters listed in Table S1. The green markers are result of a similar simulation done with reduced coherence times to match the steeper slope of the data, assuming identical relaxation and transverse relaxation times for both transmons and $T_1=T_2$ that yields slope as $3 \gamma_{l,r}^{01}/2$. The purple markers are a result of simulation including a fixed detuning on each qubit,  $\delta^{01}_l = 483$ kHz and $\delta^{01}_r = -342$ kHz. The red markers are a result of a simulation that optimizes the detunings and a small asymmetry in $g_{l,r}^{01}$ for each point to best fit the experimentally measured error-time (blue markers). In each case, $\Omega^{\rm opt}$ is set by the drive amplitude that realizes the minimum steady state error achievable in the given simulation model. (b) Values of $\delta^{01}_{l,r}$ and percentage asymmetry in $g_{r}^{01}$, extracted from fitting the experimentally measured steady state error and stabilization time as a function of normalized Rabi drive strength (red markers on left plot).}
\label{fig:comparison}
\end{figure}
In order to quantitatively understand the error-time scaling observed in the experiment and diagnose the cause of deviations from the theoretically predicted performance, we perform detailed numerical studies summarized in Fig.~\ref{fig:comparison}.
The steady-state error and time estimated from full simulations of the master equation reported in Methods fall on a straight line with slope $1/T_B$ (dashed-black), confirming the prediction from the semi-classical rate equation model.
Experimentally measured scaling of error-time, while linear, corresponds to a steeper slope though (solid-blue). A naive expectation of reduced transmon coherence could explain the steeper slope (dashed-green), but it is inconsistent with the magnitude of the minimum reported error $\varepsilon_{\infty}^{\rm min}$. The next potential cause of this discrepancy could be a drive-induced Stark shift or detuning which leads to coherent leakage out of the target state; as shown by the simulations, assuming a constant detuning leads to a qualitatively different variation between $\varepsilon_{\infty}$ and $\tau$. On the other hand, accounting for drive-dependent crosstalk both as a Stark shift on 0-1 frequencies and an imbalance in 0-1 sideband amplitudes can simultaneously describe the faster slope and minimum error observed in the experiment. We therefore conclude that this is the most likely explanation for the increased slope.
%
%
%
%